\documentclass[notitlepage,aps,prd,twocolumn,superscriptaddress,floatfix, nofootinbib]{revtex4-2}

\usepackage[utf8]{inputenc}
\pdfoutput=1

\usepackage{enumitem}
\usepackage{braket}
\usepackage{hyperref}
\usepackage{comment}
\usepackage{amsmath,amssymb,amsfonts,mathrsfs}
\usepackage{booktabs}
\usepackage{mathtools}
\usepackage{bm}
\usepackage{dcolumn}
\usepackage{graphicx}   
\usepackage{latexsym}
\usepackage[dvipsnames]{xcolor}
\usepackage{colortbl}

\usepackage[acronym]{glossaries}

\hypersetup{
  colorlinks=true
}

\definecolor{LBlue}{rgb}{0.5,0.9,0.5}
\definecolor{LCyan}{rgb}{0.6,1,1}
\definecolor{LRed}{rgb}{0.9,0.6,0.7}
\definecolor{LGreen}{rgb}{0.98, 0.93, 0.36}

\newcommand\be{\begin{equation}}
\newcommand\ba{\begin{eqnarray}}
\newcommand\ee{\end{equation}}
\newcommand\ea{\end{eqnarray}}

\newcommand{\scri}{{\cal I}}
\newcommand{\RNum}[1]{\uppercase\expandafter{\romannumeral #1\relax}}
\newcommand{\Msol}{\ensuremath{M_{\odot}}}
\newcommand{\dd}{\mathrm{d}}

\makeglossaries
\newacronym{gw}{GW}{Gravitational Wave}

\newcommand{\Loframe}{\hat{\mathbf{L}}_0}
\newcommand{\Lframe}{\hat{\mathbf{L}}}
\newcommand{\Jframe}{\hat{\mathbf{J}}}
\newcommand{\Nframe}{\hat{\mathbf{N}}}
\newcommand{\para}{\boldsymbol{\lambda}}

\def\scri{\mathscr{I}}
\def\scrip{\scri^{+}}

\newcommand{\modelfamily}[1]{\textit{#1}} 
\newcommand{\NR}{\modelfamily{NR}}
\newcommand{\SXS}{\modelfamily{SXS}}
\newcommand{\Surr}{\modelfamily{Surrogate}}
\newcommand{\Phen}{\modelfamily{Phenom}}
\newcommand{\EOB}{\modelfamily{EOB}}

\newcommand{\LALSuite}{\textit{LALSuite}}

\newif\ifcollab

\collabtrue

\ifcollab

\newcommand{\fadeout}[1]{{\small\color{black!40}#1}}
\newcommand{\st}[1]{\fadeout{#1}} 

\else

\newcommand{\fadeout}[1]{} 
\newcommand{\st}[1]{} 

\fi

\begin{document}

\newpage


\title {Testing gravitational waveforms in full General Relativity}

\author{Fabio D'Ambrosio}
\email{fabio.dambrosio@gmx.ch}
\affiliation{Institute for Theoretical Physics, ETH Zurich, Wolfgang-Pauli-Strasse 27, CH-8093 Zurich, Switzerland} 

\author{Francesco Gozzini}
\email{gozzini@thphys.uni-heidelberg.de}
\affiliation{Institute for Theoretical Physics, University of Heidelberg, Philosophenweg 16
D-69120	Heidelberg
Germany} 

\author{Lavinia Heisenberg}
\email{l.heisenberg@thphys.uni-heidelberg.de}
\affiliation{Institute for Theoretical Physics, University of Heidelberg, Philosophenweg 16
D-69120	Heidelberg
Germany} 

\author{Henri Inchausp\'e}
\email{h.inchauspe@thphys.uni-heidelberg.de}
\affiliation{Institute for Theoretical Physics, University of Heidelberg, Philosophenweg 16
D-69120	Heidelberg
Germany} 

\author{David Maibach}
\email{d.maibach@thphys.uni-heidelberg.de}
\affiliation{Institute for Theoretical Physics, University of Heidelberg, Philosophenweg 16
D-69120	Heidelberg
Germany} 

\author{Jann Zosso}
\email{zosso.jann@bluewin.ch}
\affiliation{Institute for Theoretical Physics, ETH Zurich, Wolfgang-Pauli-Strasse 27, CH-8093 Zurich, Switzerland}

\date{\today}


\begin{abstract}
We perform a comprehensive analysis of state-of-the-art waveform models, focusing on their predictions concerning kick velocity and inferred gravitational wave memory. In our investigation we assess the accuracy of waveform models using energy-momentum balance laws, which were derived in the framework of full, non-linear General Relativity. The numerical accuracy assessment is performed for precessing as well as non-precessing scenarios for models belonging to the \EOB{}, \Phen{}, and \Surr{} families. We analyze the deviations of these models from each other and from Numerical Relativity waveforms. Our analysis reveals statistically significant deviations, which we trace back to inaccuracies in modelling subdominant modes and inherent systematic errors in the chosen models. We corroborate our findings through analytical considerations regarding the mixing of harmonic modes in the computed kick velocities and inferred memories.
\end{abstract}


\pacs{98.80.Cq}

\maketitle


\newcommand{\myhyperref}[1]{\hyperref[#1]{\ref{#1}}}


\section{Introduction} 
\label{sec:intro}
The first direct detection of gravitational waves (GWs) from the merger of two black holes in 2015~\cite{PhysRevLett.116.061102} marked a milestone in the field of astrophysics, confirming a key prediction of Einstein's General Theory of Relativity (GR) and ushering in a new era of observational astronomy. 
The waveforms matched against the GW signal encode a wealth of information, including the masses and spins of the binaries, the distances to the sources, and the geometry of their motion. By analyzing these waveforms, astrophysicists can uncover the underlying physical processes, discern the properties of exotic objects, and validate theoretical models with unprecedented precision.

To date, the process of parameter estimation and the testing of GR necessitate numerical modeling of gravitational waveforms across a wide range of source parameters, including masses, spins, and other relevant factors pertaining to the merging objects.
These modelled waveforms are used as fitting templates with respect to actual data. The effectiveness of this template-to-signal match hinges on the precision of the estimated waveforms. 
Therefore, in order to extract meaningful information from signals, it is crucial to construct comprehensive and accurate waveform templates that faithfully capture the physics of GR.
This requirement is further substantiated by the expected increase in signal-to-noise-ratio of future GW observatories such as LISA~\cite{colpi_lisa_2024}, the Einstein Telescope~\cite{maggiore_science_2020}, and the Cosmic Explorer~\cite{evans_cosmic_2023}. Their increased resolution and sensitivity opens them up to more subtle effects like the gravitational memory~\cite{zeldovichRadiationGravitationalWaves1974, christodoulouNonlinearNatureGravitation1991a, thorneGravitationalwaveBurstsMemory1992, PhysRevD.44.R2945, PhysRevD.80.024002, favataNONLINEARGRAVITATIONALWAVEMEMORY2009} and they may even reveal physics beyond GR~\cite{Heisenberg:2018vsk}.

However, deviations between templates and the actual waveforms contained in the observational data introduce systematic biases, compromising the reliability of the information extracted from the signal. To counteract such biases, a diverse set of template waveforms is employed when analyzing data from GW instruments. Among these template waveforms, the ones obtained from Numerical Relativity (NR) simulations are the most reliable. Their disadvantage, however, is their consumption of vast amounts of computational resources for each simulation, i.e., for each choice of parameters describing an individual merging scenario. This time-consuming and resource-intense process poses a significant challenge, particularly as the volume of data to be processed is expected to increase drastically in the upcoming years with the advent of multiple ground- and space-based instruments like LISA~\cite{LISA_2023} and LIGO/Virgo~\cite{ALIGO_2015, AVIRGO_2014}. 
Furthermore, as the measuring precision advances, deviations from GR~\cite{Heisenberg:2018vsk} may reveal themselves in the observed data. Detecting such deviations necessitates an expanded parameter space to account for alternative descriptions of gravity, which consequently amplifies the number of waveforms against which the data must be tested.

To address these challenges, and in particular to tackle the efficiency issue of the waveform generation process, multiple alternatives to NR simulations were established. Prominent representatives of alternative waveform models are the \textit{Surrogate} models~\cite{Blackman:2017pcm,Blackman:2017dfb, NRSur7dq4_paper}, phenomenological models~\cite{PhysRevD.93.044006,PhysRevD.93.044007, PhysRevD.103.104056, IMRPhenomTPHM_paper} and effective-one-body simulations~\cite{SEOBNRv1, SEOBNRv2T, SEOBNRv3, SEOBNRv4, SEOBNRv4PHM_paper, ramosbuades2023seobnrv5phm}. 
To obtain reliable results within reasonable timescales, the models adopt distinct strategies to compute the gravitational strain. 
Each model focuses on different physical aspects of compact binary coalescence and is capable of producing a waveform within certain parameter ranges in an efficient manner. To understand what distinguishes the different approximant families, it is helpful to divide the compact binary coalescence into three phases: the inspiral, the merger, and the ringdown phase. The objective of each approximant family is to generate waveforms that replicate NR without resorting to the computationally expensive procedure of numerically solving GR in the strong-field regime. While most models follow a Post-Newtonian (PN) ansatz for the inspiral (see also section~\ref{sec:models}), merger and ringdown are simulated using various approaches, such as effective-one-body or phenomenological models, as mentioned above. After modeling each of the three phases separately, it is necessary to patch these segments together to obtain a coherent waveform for a specific event. Given that two of the three phases are modelled in fundamentally different ways, it is evident that the process of patching waveform segments together introduces errors and further discrepancies between different approximants.

In this work, we provide an extensive and detailed comparison of state-of-the-art waveform models based on their prediction of physical quantities such as the remnant's kick velocity and the inferred gravitational wave memory. We compare the models to events extracted from NR simulations, as well as between themselves for events beyond the scope of the available NR catalogue. Our analysis prioritizes non-precessing events for the evaluation of kick velocities, and we extend the analysis of the memory to include precessing as well as non-precessing events. Our primary tool for effecting these comparisons are the so-called \textit{energy-momentum balance laws}~\cite{Ashtekar:2019viz}. These exact mathematical results were derived within full, non-linear GR and provide an infinite tower of constraints for waveform models. After performing a decomposition of the balance laws into spherical harmonics, we obtain a simple constraint equation as well as equations for directly computing kick velocity and inferred memory for any given waveform model. 
A proof of concept that the balance laws can be used as diagnostic tools for assessing waveform accuracy was provided in~\cite{Khera:2020mcz,Heisenberg:2023mdz}. Here, our objective is to fully leverage the strength of these tools for a comparative analysis between different waveform models. 
Our study builds upon prior research, which concentrated exclusively on analytical models~\cite{Heisenberg:2023mdz}, the kick velocity~\cite{Borchers:2021vyw, Borchers_2023}, or gravitational memory~\cite{Khera:2020mcz, Mitman:2020bjf}. We extend these earlier investigations and reveal systematic trends in waveform templates.

Our paper is structured as follows. In section~\ref{sec:models} we review the families of the waveform generators which are the subject of our analysis in later sections. We provide a detailed account of their conventions, and we describe their generation procedures and implementations. We further compare the use of reference frames and strategies employed for handling waveform alignment and precession within each waveform family. 
Section~\ref{sec:alignmentandparas} delves into a thorough discussion of our own alignment procedures, where we explore various strategies and discuss the outcomes for each cases. In section~\ref{sec:events} we list the binary black hole (BBH) merger events that enter our comparative analysis. We categorize them into two groups: those generated from NR (referred to as ``catalogued'') and those falling outside the NR parameter space (referred to as ``non-catalogued''). Section~\ref{sec:theory} revisits the framework and formulation of the energy-momentum balance laws, with a focus on a qualitative understanding and an emphasis on their power for assessing the accuracy of waveform models. We also perform a decomposition of the balance laws into spherical harmonics in order to distill equations for kick velocity and GW memory.  

The core of our investigations is found in section~\ref{sec:analysis}, where we analyze kick velocity, memory, and the relative differences between the selected models, based on the equations derived in section~\ref{sec:theory}. Special attention is given to the impact of mode content and selection. Finally, in section~\ref{sec:discussion}, we discuss the outcomes of the model comparison, considering aspects such as model performance, alignment strategies, and mode selection.


\section{Overview of waveform models}
\label{sec:models}
We start by presenting the most recent waveform model families that have been prominent in GW research and that we will employ in this study. We establish a unified parameter space for all models, provide an overview of their modeling procedures, and discuss their strengths and weaknesses qualitatively. For a more comprehensive overview encompassing the models considered in this work, we direct readers to~\cite{lisaconsortiumwaveformworkinggroupWaveformModellingLaser2023}.

In this investigation, we focus on an ensemble of four different waveform families available through the \LALSuite{} software collaboration~\cite{ligoscientificcollaborationLALSuiteLIGOScientific2020}. 
The list of models contains four main families, which we label \NR{}, \Surr{}, \Phen{}, \EOB{} according to their generation mechanisms. Among these four families, we focus more extensively on individual representatives. Specifically, we analyze the following implementations: \SXS{} and its \LALSuite{} interface~\cite{Schmidt:2017btt}, the \Surr{} model \textit{NRSur7dq4}~\cite{NRSur7dq4_paper}, \EOB{} being represented by \textit{SEOBNRv4PHM}~\cite{SEOBNRv4PHM_paper} and \textit{IMRPhenomTPHM}~\cite{IMRPhenomTPHM_paper} as the phenomenological model. For simplicity, in the remainder of this work we shall often identify the specific model (e.g. \textit{NRSur7dq4}) with its family (e.g. \Surr{}). The numerical relativity data from the \SXS{} collaboration is tested through the available LALsuite interpolation~\cite{Schmidt:2017btt} which we refer to as \NR{}. 

The chosen families represent distinct approaches to waveform simulations. The \NR{} family represents full numerical relativity simulations, offering the most comprehensive and reliable waveforms against which the other models are competing. The alternative ``approximant'' models are based on the phenomenology of binary mergers (\Phen{}), effective-one-body simulations (\EOB{}), and interpolation of NR simulations (\Surr{}).
In Table~\ref{table:1}, we present an overview of the selected models, specifying for each the included (spin-weighted) harmonic modes and the applicable domains concerning mass-ratio $q$ and spin magnitude $|\chi_i|$.

\begin{center}
\begin{table*}[t]
\begin{tabular}{
    |p{3.4cm}|p{3.2cm}|p{2cm}|p{2.8cm}|p{1.8cm}|p{1.8cm}|
}
 \hline \bfseries
 Family & \bfseries Implementation & \bfseries Branch & \bfseries Mode Content & \bfseries $q$-range & \bfseries $|\chi_i|$-range
 \\\hline\hline 
 Numerical Relativity & \SXS{} (NRhdf5)   & precessing    &$\{(\ell,m)| \ell \leq 8\}$  &   $\lesssim 4$ & $\lesssim 0.8$ 
 \\\hline
   Surrogate & NRSur7dq4&   precessing  & $\{(\ell,m)| \ell \leq 4\}$  &$\leq 4$&$\leq 0.8$
   \\\hline
  Effective One Body & SEOBNRv4PHM & precessing & $\{(\ell,m)| \ell \leq 5\}$&  -&- 
  \\\hline
 Phenomenological & IMRPhenomTPHM  & precessing & $\{(\ell,m)| \ell \leq 5\}$ & -&- 
 \\\hline
\end{tabular}
 \caption{
    Summary of the families, specific implementations, mode content and parameter space of the waveform models considered in this work.}
    \label{table:1} 
\end{table*}
\end{center}

\subsection{Waveform families}
Generally speaking, the precessing BBH problem for quasicircular orbits is parametrized by seven parameters: the mass ratio $q = m_1/m_2 \geq 1$ and two spin vectors $\chi_{1,2}$, each containing three independent components. Here, the index 1 (2) refers to the heavier (lighter) BH. In the following, we abbreviate all parameters relevant to a BBH merger using the vector $\para$. 
For the non-precessing case, the parameter space reduces to a total of only three parameters as the spins $\chi_{1,2}$ are aligned (or anti-aligned) and only their magnitude is relevant for the calculation of the strain. 

Using this common parameter vector, in what follows, we briefly outline the general working methods of the waveform families and highlight certain steps that are essential to identify strengths and shortcomings of each model on a phenomenological level. 

\paragraph*{Numerical Relativity:} The bedrock of waveform modelling is formed by numerical simulations of the non-linear Einstein field equations, which until today are the only means of obtaining accurate waveforms for the strong gravity regime of a BBH coalescence. The publicly available simulations are distributed across various catalogues, with one of the most comprehensive being the catalogue produced by the SXS collaboration~\cite{boyleSXSCollaborationCatalog2019}. The NR code employed in these simulation is the Spectral Einstein Code (SpEC)~\cite{kidderBlackHoleEvolution2000}. The distinguishing feature of SpEC is to utilize so-called \emph{spectral methods} for the numerical approximation of the derivative operators in the relevant differential equations at hand. In essence, while the conventional approach for approximating the derivative of a function on a given grid involves variations of the finite-difference method, spectral methods instead consider the expansion of the function to be differentiated into a finite basis of orthogonal functions on the grid.

Every numerical simulation begins by generating initial data that adheres to the constraint equations of the system, with the initial parameters $\para$ serving as input. The evolution typically commences with an initial burst of spurious or ``junk'' radiation before the systems settle into a stable inspiral phase. This initial junk radiation constitutes a primary source of numerical errors in the \SXS{} waveforms, as varying resolutions in the simulations may lead to distinct initial bursts and consequently result in slightly different estimated parameters during subsequent evolution.
The simulation progresses until a shared apparent horizon forms, indicating the onset of the coalescence phase. At this juncture, the simulation is momentarily halted, and the data are interpolated onto a new grid tailored to a single extraction boundary. Subsequently, the evolution recommences through the coalescence and ringdown phases until all ringdown gravitational waves are emitted. For the resulting waveforms, the numerical error estimation involves a comparison of waveforms generated at different resolutions, specifically the highest and the next-to-highest. 

The NR waveforms obtained via the described procedure and collected in the main catalogue considered in this study do not exhibit GW memory effects. According to~\cite{mitmanComputationDisplacementSpin2020}, this is attributed to the extrapolation procedure of finite-grid data to null infinity, which utilizes polynomial interpolation. The more accurate extrapolation method based on Cauchy-characteristic extraction (CCE)~\cite{taylorComparingGravitationalWaveform2013, mitmanComputationDisplacementSpin2020} has, so far, only been implemented in a few events within the public catalogue but contains gravitational memory in the catalogued strain modes.

\subsubsection{Effective One Body}
The Effective One Body (\EOB{})~\cite{Damour_2014, Buonanno_2006} models attempt to approximate the complete waveform of binary mergers by replacing the two-body dynamics by an effective one-body description, i.e., by solving the problem of a single particle of reduced mass immersed in an ``effective'' background metric, $g^{\text{eff}}_{\mu \nu} (x^{\lambda})$~\cite{SEOBNRv1}. Solving the equations resulting from the effective background metric, one obtains the full waveform for the complete BBH merger process.

There have been several versions of \EOB{} models over the past two decades.
Arguably, the most significant advancements manifested with the introduction of the NR-enhanced models known as the Spinning Effective One Body Numerical Relativity models (\textit{SEOBNR}). These models leverage NR simulations to calibrate the free parameters intrinsic to the \EOB{} waveform.
For instance, \textit{SEOBNR} improves and calibrates its waveforms by requiring the amplitudes, time derivatives of the amplitudes, GW frequencies, and time derivatives of the GW frequencies to coincide with that of the NR simulations at the matching times. Given that NR simulations only cover discrete values of the parameter space, interpolation is employed to bridge the gaps.

Naturally, the accuracy of interpolation is limited and hence the obtained \textit{SEOBNR} waveforms span only across a constrained region of parameter space. 
To extend \textit{SEOBNR}'s coverage across a broader region, BH perturbation theory is utilized to generate waveforms for mass ratios significantly smaller than one. Combining all techniques and latest advancements, \textit{SEOBNR} constitutes a waveform model that, in principle, covers any mass ratio and nearly any spin configuration.

In general, several other adjustable parameters may be incorporated into \EOB{}'s generation pipeline, depending on whether the model accounts for the spins of the BBH constituents and whether it describes the precession of the BBH system. For instance, \textit{SEOBNRv1-v4} models describe spinning non-precessing BBHs, \textit{SEOBNRv4HM} includes higher modes in the waveform, and \textit{SEOBNRv4PHM} incorporates both higher modes and precession effects~\cite{SEOBNRv1, SEOBNRv2T, SEOBNRv3, SEOBNRv4, SEOBNRv4PHM_paper}. Notably, the latest iteration of this family, \textit{SEOBNRv5PHM}~\cite{ramosbuades2023seobnrv5phm}, is not included in this study.

\subsubsection{Phenom}
Phenomenological models (\textit{Phenom}) attempt to approximate waveforms derived through alternative methods and interpolate within the parameter space of BBH mergers. A key objective is to rapidly generate these waveforms, enabling their utilization in matched filtering searches and for parameter estimation. Here we focus on the \textit{IMRPhenom}-family of phenomenological models. In particular, we consider \textit{IMRPhenomD}~\cite{PhysRevD.93.044006,PhysRevD.93.044007} and the most recent versions \textit{IMRPhenomXPHM}~\cite{PhysRevD.103.104056} and \textit{IMRPhemonTPHM}~\cite{IMRPhenomTPHM_paper} models. The latter two mainly differ by the domain in which the waveform is generated, where \textit{X} refers to the frequency domain and \textit{T} to the time domain.

The starting point of the \textit{IMRPhenomD} model is the decomposition of the $\ell = |m| = 2$ component of a NR waveform in the frequency domain as $\tilde{h}_{22}(f) = A(f) \, e^{- i \phi(f)}$.
The functions $A(f)$ and $\phi(f)$ can be approximated by specifically designed parameterized functions which yield a good fit across a diverse range of binary parameters. Within the \textit{IMRPhenomD} model, these functions are piecewise smooth and effectively capture the inspiral, intermediate, and merger-ringdown regions. 
For the inspiral region, an augmented \EOB{} waveform is employed. The augmentation involves incorporating a sum of five parametrized powers of frequency for the phase, and three parametrized powers of frequency for the amplitude. These supplementary terms facilitate a more precise fitting of waveforms. The fitting functions for the intermediate and merger-ringdown regions are entirely phenomenological. They consist of parameterized polynomials as well as more general functions to capture specific features of the waveform. 

In the calibration step, each phenomenological waveform parameter $\boldsymbol \Lambda$ (17 in total) is allowed to depend on the binary parameters $\para$ (11 in total) that went into the generation of the NR waveforms. This greatly improves the flexibility and accuracy of the model, which can now depend on as many as $17 \times 11 = 187$ free parameters for the fit (listed in Appendix C of~\cite{PhysRevD.93.044007}).
Other optimizations that have been implemented in subsequent versions include additional parameters in the fundamental $h_{2,2}$ mode, the introduction of parameters for higher modes, and the inclusion of NR waveforms of higher mass ratios.

Two additional aspects worth noting are time-domain phenomenological models and precession. Conceptually, the only distinction for time-domain models is that instead of starting with the frequency domain waveform~$\tilde{h}_{2,2}(f)$, the process initiates with the time-domain waveform~$h_{2,2}(t)$.
One then proceeds as for the \textit{IMRPhenomD} model by decomposing it into an amplitude and phase function. Time-domain models are part of the \textit{IMRPhenomT*} series of models, such as \textit{IMRPhenomTHM}~\cite{PhysRevD.105.084039}. Compatibility with precessing events has been included in more recent models by ``twisting up'' non-precessing waveforms~\cite{PhysRevLett.113.151101}. This procedure distributes the power across different $m$-modes for a fixed $\ell$ of the strain's mode decomposition based on the precession of the individual spin vectors of the BH's involved in the merger.
In this work, we solely consider the time-domain precessing model \textit{IMRPhemonTPHM}~\cite{IMRPhenomTPHM_paper}.

\subsubsection{Surrogate}
A surrogate waveform model utilizes a collection of precomputed waveforms across a defined set of parameters and subsequently performs an interpolation in the parameter space between these waveforms. The objective is to rapidly generate a waveform for any given parameter values. A surrogate waveform can be computed much more rapidly than the underlying (typically NR) calculation, and its accuracy can match that of the underlying model provided there is a sufficiently extensive set of precomputed waveforms covering a broad range of parameter space.

One can broadly distinguish between ``pure'' surrogate models and ``hybrid'' surrogate models. Pure surrogate models~\cite{Blackman:2017pcm,Blackman:2017dfb, NRSur7dq4_paper} take a set of NR waveforms (e.g. 866 different simulated events for~\cite{NRSur7dq4_paper}) close to the desired point in parameter space as input and employ parametric fits to extrapolate the waveform across a continuous set of parameters. The extrapolation involves several assumptions, with the alignment of the NR waveforms considered for a given simulation being a critical factor. In general, each NR waveform must be associated with a single value of the parameter vector $\para$, even though some parameters (especially spin directions) are time-dependent. 
This procedure involves various subtleties, one of which is the alignment procedure further elaborated upon in section~\ref{sec:alignmentandparas}. 

In order to address some of the major shortcomings of numerical simulations --- namely, a low number of inspiral orbits and a narrow range in parameter space, especially in the mass ratio --- so-called \emph{hybrid} models have been developed. Hybrid surrogate models~\cite{Varma_2019:hyb} construct a surrogate waveform by treating the inspiral independently from the NR waveforms with respect to which the model interpolates, using the same masses and spin components. Each full waveform consists of a semi-analytical waveform at early times (up to roughly $20$ orbits before the ringdown) that is smoothly attached to a NR waveform at late times which is constructed as for the pure surrogate models. For the early inspiral, a mixture of PN and EOB calculations is used. 
The PN waveform corresponding to the given parameters~$\para$ is improved using the \textit{SEOBNRv4} model~\cite{SEOBNRv4}. This results in a EOB-corrected waveform that remains faithful to the NR waveform until much later times, whilst the pure PN waveform becomes inaccurate.
Note that since one restricts simulations to the three-dimensional space of non-precessing BBHs for hybrid surrogate models, fewer simulations are necessary compared to the seven-dimensional precessing case.
The resulting hybrid  model is an NR-based surrogate model to span the entire LIGO frequency band for stellar mass binaries; assuming a low-frequency detector cutoff of $20$~Hz, this model is valid for total masses as low as $2.25\, M_\odot$. It is based on $104$ NR waveforms in the range given in Table~\ref{table:1}. For the practical purpose of including precessing events in our analysis, however, we focus on the pure \textit{NRSur7d4}~\cite{NRSur7dq4_paper} model\footnote{Note that recently, the first \Surr{} model with memory effects, obtained by training the model with CCE-extrapolated numerical simulations~\cite{mitmanComputationDisplacementSpin2020}, has been published~\cite{yooNumericalRelativitySurrogate2023}.}.

\subsection{Common difficulties of waveform models}
\subsubsection{Reference frames}
In the process of calculating waveforms, each model uses multiple reference frames which are best suited for their individual computations.
While the waveform families are generally different, there is an overlap in some commonly used frames that simplify numerical calculations.
There are essentially four commonly used frames reoccurring in the algorithms that process the waveforms.
Following the notation in~\cite{Schmidt:2017btt}, these are:
\begin{description}[style=unboxed,leftmargin=0cm]
    \item[$\Loframe$- or source-frame] This frame uses the orbital momentum vector $\vec{L}$ at a certain reference frequency $f_\text{ref}$ (or time $t_\text{ref}$) to define the $z$-axis of a right-handed Cartesian coordinate system. 
    This is a time-independent frame. 
    \item[$\Jframe$- or inertial-frame] Here the total angular momentum vector $\vec{J}$ is used to define the $z$-axis of a right-handed Cartesian coordinate system. This is another time-independent frame.
    \item[$\Nframe$- or observer-frame] This frame's $z$-axis is defined by the line-of-sight vector $\hat{N}$ from the observer to the centre-of-mass of the binary. This frame is also time-independent.
    \item[$\Lframe$- or coprecessing-frame] This frame coincides with the $\Loframe$ frame when the orbital frequency $f$ (or time $t$) is equal to $f_\text{ref}$ (or $t_\text{ref}$). This is a time-dependent frame. In fact, it is obtained by evolving the $\Loframe$ in time and it is used to keep track of the precession of the orbital plane. With a small modification of the coprecessing frame, one obtains the so-called \textit{coorbital-frame}, in which the BHs are always on the $x$-axis, with the heavier BH on the positive half.
\end{description}
While these descriptions established a point of origin and one axis of a Cartesian coordinate system, a full definition of the different frames is yet to be provided, i.e., we are left with rotational freedom around the defined axis. One can either constrain this freedom through a more complete definition of frames or leave the remaining freedom to be minimized through an alignment of waveforms. In this work, we adopt the latter approach, as detailed in section~\ref{sec:alignmentandparas}.

\subsubsection{Alignment:}
All the approximants under consideration encounter the challenge of comparing waveforms originating from different sources, each potentially using a distinct coordinate system.
To perform an alignment of different waveforms, one aims to identify an initial state in the inspiral from which on each model evolves the BBH forward in time. If this initial state is found, there is a unique rotation translating one frame into each other. 

A common strategy is to first identify the time of coalescence for each waveform by considering the peak time of the strain amplitude, and set it to $t = 0$. Then, when comparing to NR simulations, it is possible to obtain the full state of the binary system at some fixed earlier time $t_\text{ref} < 0$, sufficiently deep in the inspiral phase (but not too early, to avoid incurring in so-called ``junk radiation''). Subsequently, the unique rotation of the system that puts the two black holes in the desired configuration in the chosen reference frame, for example with the more massive black hole along the positive $x$-axis, can be determined. An instance of this procedure for precessing \Surr{} models is described in~\cite{Blackman:2017dfb}. There, at reference time $t_\text{ref} = -4500M$ the data from the numerical simulations are used to find the direction of the orbital angular momentum of the binary, which define the instantaneous orbital plane at $t_\text{ref}$. A unique rotation and a small error correction are then applied to move the more massive black hole on the $x$-axis and fix any residual ambiguity of the frame definition mentioned above. Note that this implies that the ``initial'' parameters $\para$ that define the waveform are set at $t_\text{ref}$ only \emph{after} the alignment. 
 However, for the purpose of model comparison, it is necessary to obtain waveforms corresponding to the \emph{same} initial conditions at the same initial time. One possible approach is to perform numerical optimization over the binary parameters at a specific fixed reference time. This strategy has been employed for benchmarking \emph{IMRPhenomXPHM}~\cite{Pratten:2020b}, and we also adopt this approach in our study. We refer to section~\ref{sec:alignmentandparas} for a full discussion of the subtleties involved in this choice of the alignment procedure as well as the practical implementation.

\subsubsection{Precession}
\label{subsubsec:precession}
Generating waveforms for precessing events, each waveform model family has evolved different strategies.
Recent \Phen{} models describe precession effects via the so-called twisting up of the waveform~\cite{PhysRevLett.113.151101}. This procedure is applied to a waveform obtained for the corresponding non-precessing BBH merger. Assuming that the precession frequency is much lower than the orbital frequency, one can employ a stationary-phase approximation and obtain a map between precessing and non-precessing BBH mergers. 
In this regime, the predominant impact of precession manifests as amplitude and phase modulation, expressible through a time-dependent rotation of the non-precessing system. Once the specific form of this rotation and its dependence on the binary parameters are established, it can be directly applied to the waveforms. Such a time-dependent rotation naturally mixes different modes. In particular, if the non-precessing waveform contains any mode of a certain $\ell_0$, then the precessing waveform receives contributions from \emph{all} modes up to that $\ell_0$. As such, all precessing \Phen{} models contain strain modes of the form $\{(\ell,m) \mid \ell \le \ell_0\}$ for some $\ell_0$, even if the underlying non-precessing model does not include modes for all $m$ within that range. Crucially, the absent modes in the underlying waveform do not remain absent in the precessing waveform, but the lacking information is now distributed across all modes of that specific $\ell_0$. In such cases, none of the modes of that $\ell_0$ can be expected to be accurately modelled. For instance, the \textit{IMRPhenomXHM} model includes the~$h_{4,4}$ mode in $\ell=4$, hence the twisted up \textit{IMRPhenomXPHM} model~\cite{Pratten:2020b} receives contributions from all $(4,m)$ modes with $|m| \le 4$ for a precessing binary, but none of them is ``complete'' due to lack of, e.g., the $(4,3)$ mode in the \textit{IMRPhenomXHM} model. 

Similar considerations apply to the \EOB{} model \textit{SEOBNRv4PHM}. The latter is an augmented version of the non-precessing model \textit{SEOBNRv4HM} (hence sharing effectively the same mode content), where precession is introduced by projecting the spin of the effective background Kerr spacetime onto the orbital angular momentum~\cite{SEOBNRv4PHM_paper}. 

On the contrary, precessing \Surr{} models adopt a different approach that refrains from making specific assumptions about the modelling of the underlying waveform~\cite{Blackman:2017dfb}. In principle, they contain the full mode content up to the claimed value of $\ell$ for precessing as well as non-precessing BBH mergers, where each mode is subject to the uncertainties introduced in the fitting procedure with NR waveforms. Consequently, we anticipate superior performance from these models compared to other approximants when precession is a factor. Our analysis, as detailed in section~\ref{sec:analysis}, confirms this expectation.

\subsection{Strengths and weaknesses}
Fully numerical simulations can be regarded as the most accurate means to obtain BBH waveforms. Except for the early inspiral, where PN approximations are highly accurate, the intricate equations governing coalescence can only be comprehensively solved through numerical simulations. These sophisticated computations are resource-intensive, demanding substantial computational power. Consequently, only a finite number of discrete points in parameter space have been simulated and compiled in catalogs thus far.
The sparse parameter space coverage of the \SXS{} catalogue~\cite{boyleSXSCollaborationCatalog2019}, which is still the most comprehensive one, motivated alternative GW models.

As previously explained, the \Surr{} models utilize a specific set of NR waveforms to parametrically fit a new waveform located in their vicinity in parameter space. However, this implies that \Surr{} models are once again confined to a specific region of parameter space, albeit to a significantly lesser extent than when only considering the available catalogs.

For \EOB{} models, the restrictions in parameter space are weaker than for \Surr{} models. Due to the analyticity and adjustability of the \EOB{} models, they can cover a large portion of the BBH parameter space, once the calibration is performed against NR simulations. For instance, \textit{SEOBNRv1} covers arbitrary mass ratios and black hole spins in the range $-1\leq \chi_c \leq 0.7$ for the dominant $h_{2,2}$ mode, with further improvements in later versions. 

\Phen{} models, on the other hand, again rely predominantly on the calibration to NR simulations. However, as they produce only an approximate waveform these models are still very fast and can be easily extended by introducing more parameters and different fitting functions. The precision can be increased by fitting to more NR simulations, which will naturally yield a drop in performance.

As one might anticipate, the expansion in parameter space comes with certain trade-offs. For instance, in the case of \Surr{}, the non-precessing waveforms encompass only a limited number of orbits before the merger. As mentioned, this limitation is addressed in hybrid \Surr{} models to attain an accuracy comparable to \EOB{} for the inspiral segment. In extensive data analysis scenarios, however, the evaluation of \EOB{} models can be computationally expensive, potentially undermining the purpose of being an approximant.

In summary, each model has its ``region of proficiency'' and delivers a reliable, accurate waveform for a given point in parameter space.  Although the region of applicability may be smaller for some models, they might excel in precision in other aspects. The choice of the waveform model should be based on the specific requirements at hand, considering factors such as relevant mode content, the segment of the waveform of interest, the importance of quality versus quantity, and so on.


\section{Waveform Alignment}
\label{sec:alignmentandparas}
In what follows, we delineate the alignment procedure employed in this work, which, in some aspects, shares similarities with the previously presented approaches of the waveform models. It constitutes the first step towards a comprehensive comparison of different waveform models.

\subsection{Alignment and residual ambiguities}
\label{subsec:alignment}
To ensure a meaningful comparison, the output waveforms computed by each  model must be translated into a common frame. This task has been primarily handled by the \LALSuite{} framework, which encompasses the relevant approximants and many more. The waveforms generated through \LALSuite{} are already aligned in a common frame, up to a rotation in the orbital plane, or equivalently, up to a phase factor denoted hereafter as $\phi_\text{ref}$. Provided the waveforms of the approximants are of equal length (i.e., start at a common reference frequency), the orbital phase remains the only obstacle in the alignment of non-precessing waveforms.

An other issue arises when waveform models are compared to the \SXS{} catalogue: For a given event associated with a specific set of spins and initial masses, it is possible that the maximal waveform domain of each approximant varies due to different restrictions on the reference frequency $f_\text{ref}$\footnote{Here, we refer to the reference frequency as the initial time step from which on the generated waveform is non-trivial. This frequency can be translated to the number of orbits included before the merger, which generally varies throughout waveform models.}. In this case, we re-generate the corresponding events for each approximant using a new minimal common $f_\text{ref}$. While the \LALSuite{} approximants can produce waveforms with different initial or reference frequencies $f_\text{ref}$ for a given event (within intrinsic constraints, see Table~\ref{table:1}), the \SXS{} waveforms come with a fixed reference frequency that cannot be altered. As a result, the necessity of adjusting NR waveforms ``by hand'' arises, which in turn requires an additional alignment step with respect to the phase factor $\phi_\text{ref}$. 

For precessing events, analogous arguments apply. Here, the ambiguity to be resolved by the alignment extends to the initial directions of the individual spin vectors of the black holes, introducing four more independent parameters (the angles subtended by the spin vectors) to the alignment process, in addition to $\phi_\text{ref}$.

Both the adjustments of $\phi_\text{ref}$ and the initial spin vectors are achieved through minimizing a mismatch function $\mathcal{M}(\phi_\text{ref}, \Omega_1, \Omega_2)$, where $\Omega_i = (\phi_{\chi_i}, \theta_{\chi_i})$ are the pairs of angles for each spin vector.
Once minimization is accomplished, we posit that any discrepancies identified when comparing physical quantities on the aligned waveforms can be attributed to inherently distinct evaluations of the binary configuration for each individual model\footnote{Naturally, this is a crude assumption. However, an in-depth analysis of potential shortcomings in the alignment procedure beyond what is discussed in this section lays out of the scope of this work.}. 

The mismatch function requires an input that is related to the gravitational strain. Our primary strategy is to align with respect to the dominant $h_{2,\pm2}$ mode, which corresponds to choosing as mismatch function
\begin{equation}\label{equ:mismatch}
    \mathcal{M} (\phi_\text{ref}, \Omega_1, \Omega_2) \coloneqq 1 - \frac{\langle\tilde{h}^\text{ref}_{2,2}, \tilde{h}^\text{align}_{2,2}\rangle}{\lVert\tilde{h}^\text{ref}_{2,2}\rVert \,\lVert\tilde{h}^\text{align}_{2,2}\rVert} \,.
\end{equation}
Here, $\tilde{h}_{2,2}$ corresponds to the Fourier transformed $(\ell = m = 2)$-mode of the strain in the orthogonal decomposition
\begin{align}\label{equ:full_strain_decomposition}
    h = \sum_{\ell ,m} h_{\ell ,m}(u) _{-2}Y_{\ell ,m}(\theta, \phi)\, ,
\end{align}
where $_{-2}Y_{\ell ,m}$ are the spin-weighted spherical harmonics of weight $-2$. The inner product $\langle\cdot,\cdot\rangle$ is defined in the frequency domain as
\begin{equation}\label{eq:M-def}
\langle \tilde h_1, \tilde h_2 \rangle \coloneqq \int \dd f\, \tilde h_1(f)\, \tilde h^*_2(f)\,,
\end{equation}
where $\tilde{h}^*$ is the complex conjugate of $\tilde{h}$. Naturally, this inner product induces a norm, which we denote as $\lVert\cdot\rVert$. The modes $\tilde h_{2,2}^\text{ref}$ and $\tilde h_{2,2}^\text{align}$ are extracted from the reference model and the model to be aligned, respectively. Throughout the analysis, we adopt \SXS{} waveforms as reference. For regions in parameter space where no \SXS{} waveforms are  available, we take \Surr{} as reference instead.

Generally, the mismatch can be computed with respect to any strain mode or the full gravitational strain as in equation~\eqref{equ:full_strain_decomposition}. It is established common practice to align with respect to the dominant $h_{2,2}$ mode in frequency domain. 
However, there is a caveat to this approach. When minimizing $\mathcal{M}$ for the $h_{2,2}$ mode with respect to the complete Fourier-space waveform (i.e., its maximal available extent in Fourier space), the resulting alignment can obscure systematic differences between the models. This is a direct result of the procedure of minimizing over all distinctive features of the individual waveforms, including the ones relevant to qualitative comparisons of the approximants, i.e., different dynamics during the merger and ringdown stage. Since our primary interest is to align the initial sections of the waveforms, i.e., initial frequency, phase, and spins, it is more appropriate to restrict the integration of the inner product~\eqref{eq:M-def} to the inspiral phase only. Additional support for this alignment strategy can be given by noticing that the inspiral is almost exclusively modeled via a PN approximation throughout all approximants. Thus, intrinsic differences of waveform models in this frequency range can be expected to be minimal. 

In the following subsection we scrutinize this alignment strategy and determine the most suitable alignment procedure, which will be employed throughout the rest of the article.

\subsection{Testing alignment methods}\label{subsec:pref_align}
\begin{figure}
	\centering
	\includegraphics[width=0.99\linewidth]{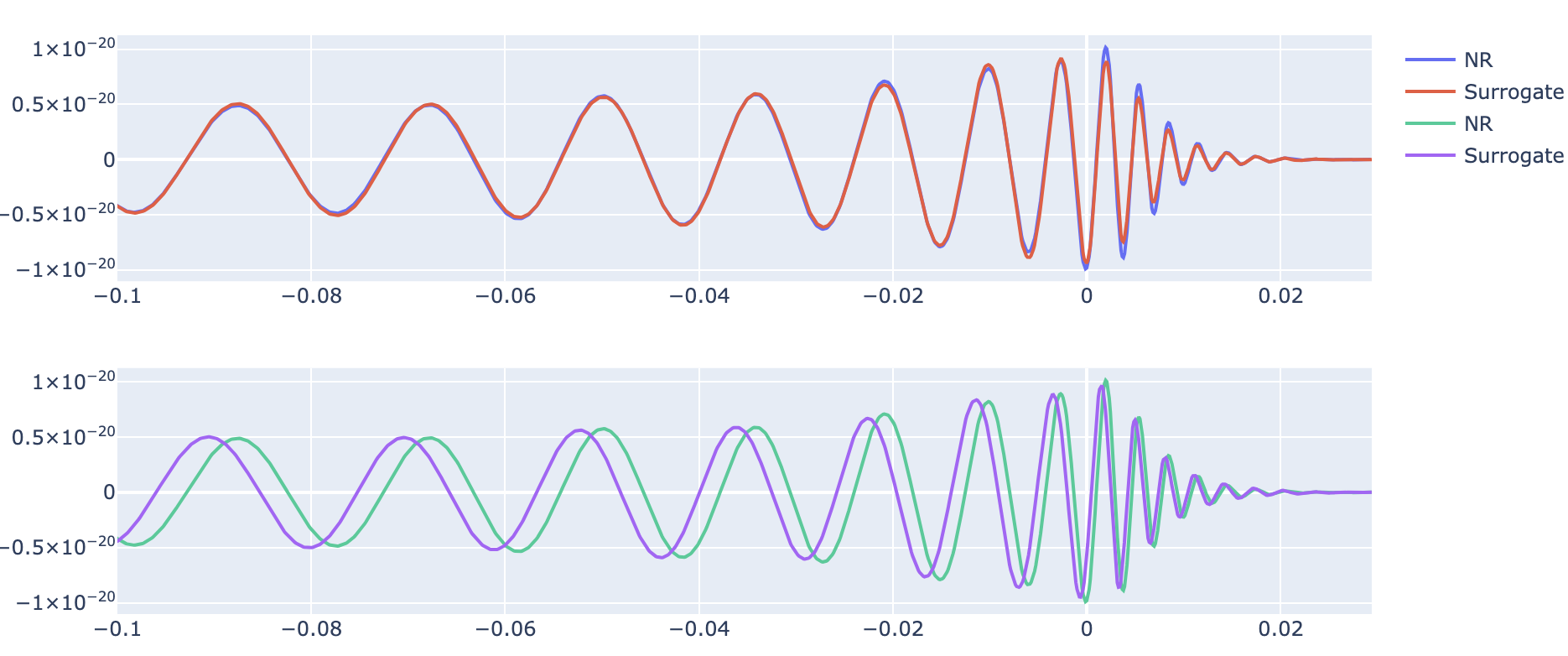}
	\caption{Aligned (top) and non-aligned (bottom) $h_{2,2}$ waveform mode for a precessing event (SXS:BBH:$1011$).}
	\label{fig:alignI}
\end{figure}

\begin{figure}
	\centering
	\includegraphics[width=0.99\linewidth]{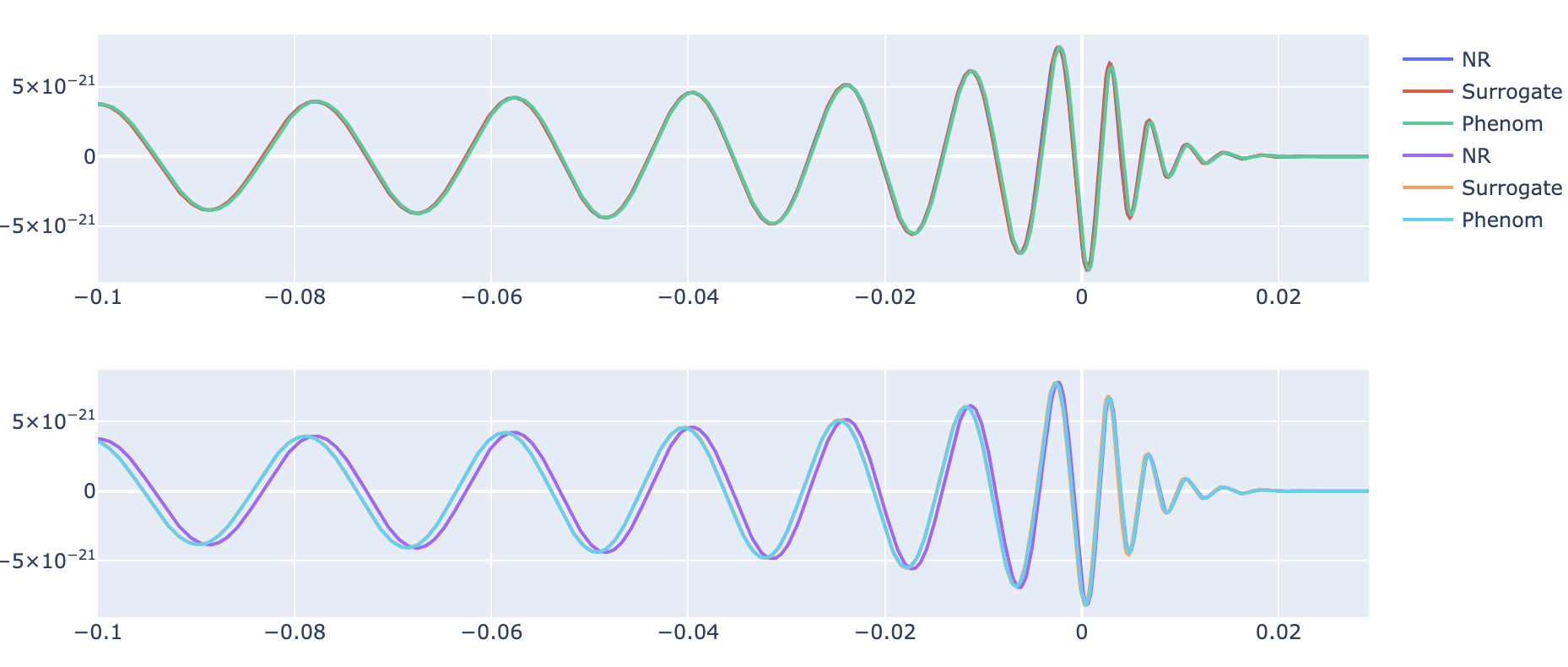}
	\caption{Aligned (top) and non-aligned (bottom) $h_{2,2}$ waveform mode for a non-precessing event (SXS:BBH:$0374$).}
	\label{fig:alignII}
\end{figure}

For the following comparison we use the  mismatch function~\eqref{equ:mismatch} but allow the use of shorter integration intervals in frequency space. An application of the standard mismatch formula~\eqref{equ:mismatch} (i.e., where the integral ranges over the full frequency spectrum of $h_{2,2}$) is displayed in Figures~\ref{fig:alignI} and \ref{fig:alignII}, for a non-precessing and a precessing waveform. In this case, the two waveforms from \SXS{} and \Surr{} demonstrate excellent agreement after the alignment procedure. The non-precessing waveform is perfectly aligned after optimization over the orbital phase parameter. The precessing event shows very good agreement in phase but some noticeable difference in the amplitude of the peaks close to the merger. 

\begin{figure}
	\centering
	\includegraphics[width=0.99\linewidth]{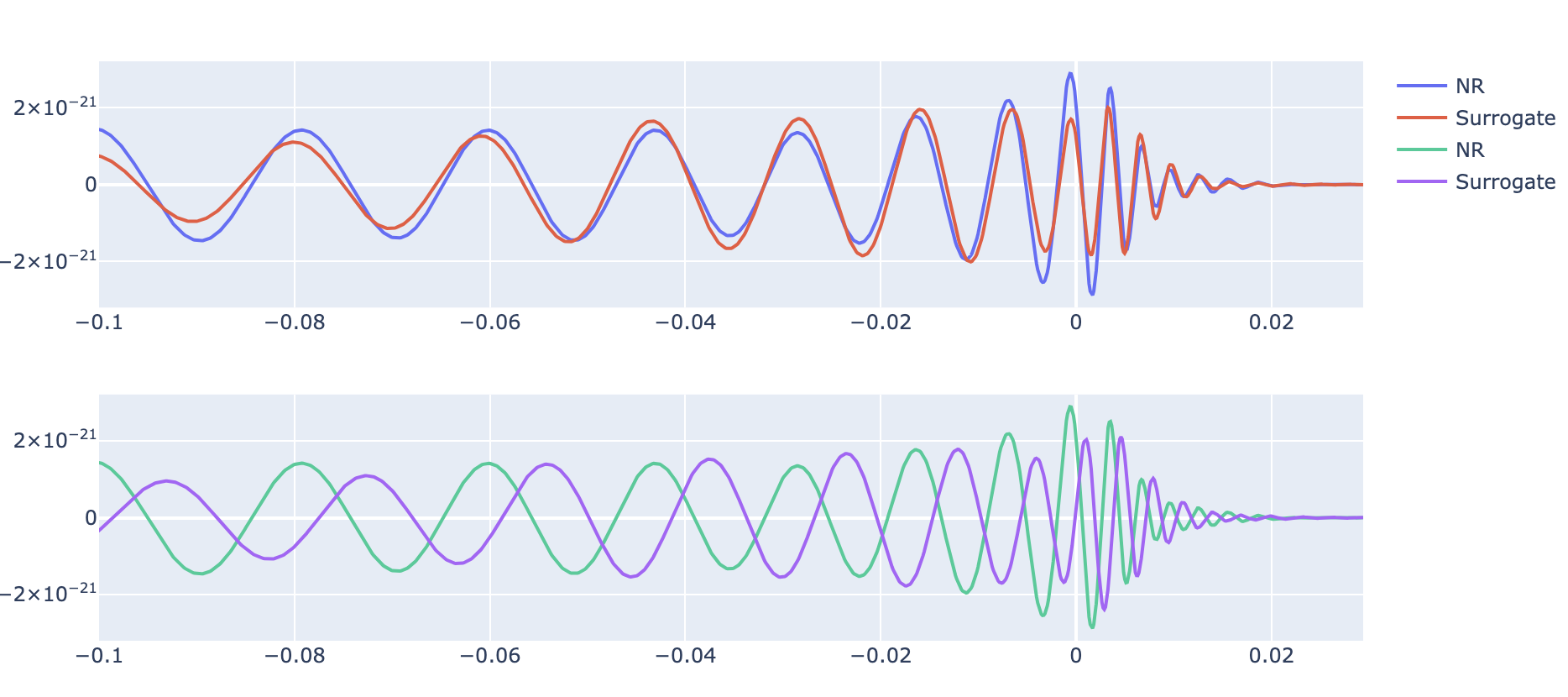}
	\caption{Aligned (top) and non-aligned (bottom) $h_{2,1}$ waveform mode for a precessing event (SXS:BBH:$1011$).}
	\label{fig:alignIII}
\end{figure}

\begin{figure}
	\centering
	\includegraphics[width=0.99\linewidth]{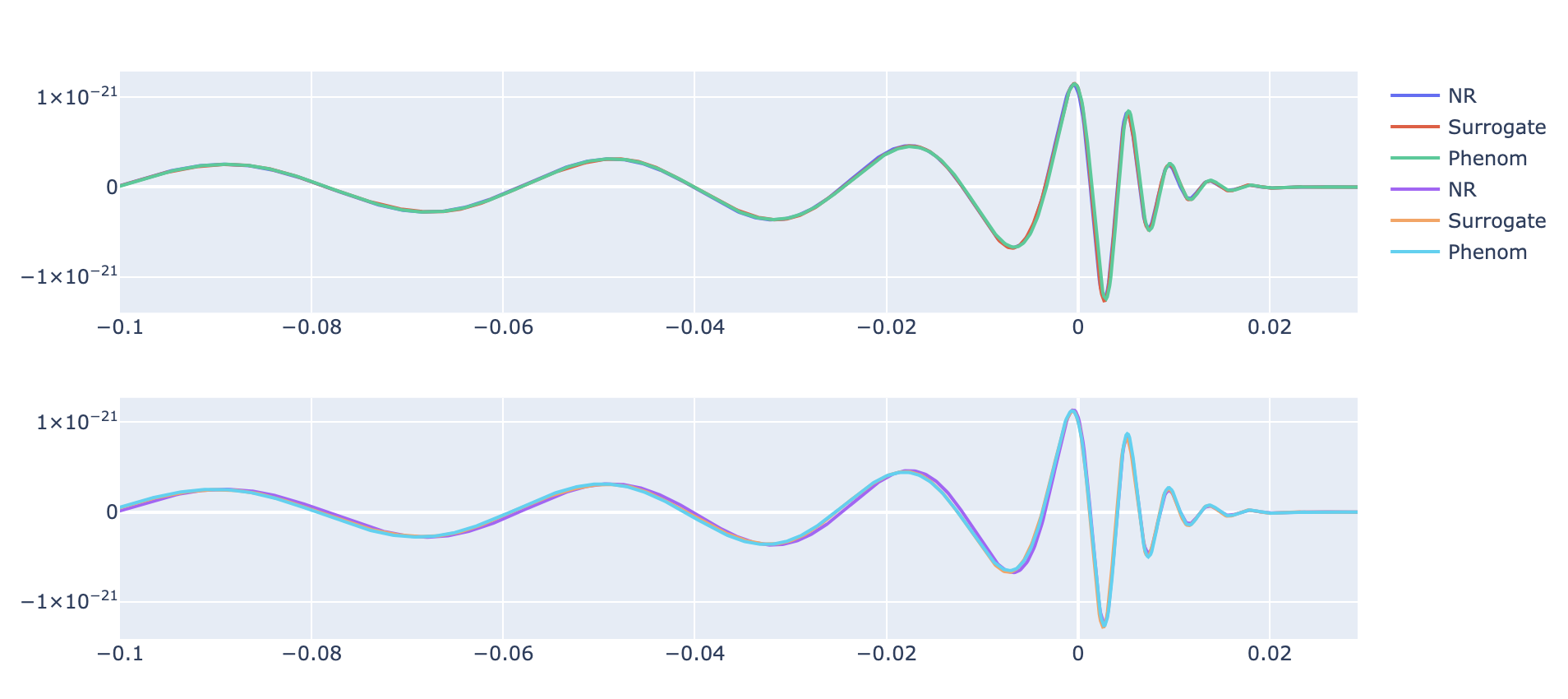}
	\caption{Aligned (top) and non-aligned (bottom) $h_{2,1}$ waveform mode for a non-precessing event (SXS:BBH:$0374$).}
	\label{fig:alignIV}
\end{figure}

The limitations of the standard alignment procedure become evident when  subdominant modes are compared pre- and post-alignment. Figures~\ref{fig:alignIII} and \ref{fig:alignIV} show the $h_{2,1}$ modes for precessing and non-precessing events prior to and after the standard alignment procedure. The aligned $h_{2,1}$ modes in the precessing case display pronounced differences, even in the inspiral phase, while in the non-precessing case there are no noticeable differences.
This simple comparison illustrates multiple issues arising in the alignment procedure. First of all, aligning with respect to the $h_{2,2}$ mode does not guarantee a perfect alignment of the remaining, subdominant modes, in particular for the precessing case. As the $h_{2,2}$ mode is the dominant one, however, the mismatch in the other modes is not our primary concern. Furthermore, while the overall mode content varies throughout waveform models, the inclusion of the $h_{2,\pm2}$ mode is guaranteed in any case, making it the preferred choice for an alignment strategy. 
Nonetheless, to address the poor alignment of subdominant modes, aligning with respect to the full strain could present a viable solution. Due to the dominance of $h_{2,2}$ within the full strain $h$, aligning with respect to the $h$, rather than with respect to solely $h_{2,2}$, results only in minor differences. In a limited test including a small sample of non-precessing events, we found that aligning with respect to the full strain results in slightly larger mismatch and differences with respect to NR regarding the quantities studied in section~\ref{sec:analysis}, i.e., the remnant velocity magnitude and direction, compared to standard alignment with $h_{2,2}$. 
This might be partially due to the models' primary focus on generating a maximally precise dominant strain mode component, rendering the subdominant modes intrinsically less exact.

Moreover, Figures~\ref{fig:alignI} and~\ref{fig:alignIII} illustrate that larger differences in the waveforms predominantly appear in the merger and ringdown phases.
As mentioned, this raises the question whether aligning with respect to only the inspiral phase suffices and whether it is more transparent regarding intrinsic model differences in the merger and ringdown phases. 
In practice, when working with time-domain waveforms, this can be tested as follows: One identifies the merger time as the peak of the strain norm and then truncates the template waveforms entering equation~\eqref{equ:mismatch} (before taking the Fourier transform) at a time well before the merger, deep in the inspiral phase. Here, we choose to cut at two-thirds of the waveform's full-time interval. This specific cutoff time corresponds to a cut-off frequency which enters the frequency domain integral in~\eqref{eq:M-def}. We compare the alignment with respect to $h_{2,2}$ with and without temporal (frequency) cut-off, over a sample of randomly chosen non-precessing and precessing events. We find that the presence of the cut-off in the alignment procedure marginally worsens the mismatch of full waveform compared to the full mismatch computation. Anticipating the analysis in section~\ref{sec:analysis}, we also find that not only the mismatch, but also the relative error in the kick velocity and direction increase as well, albeit only marginally. This result indicates that when aligning with respect to the full waveform, (intrinsic) differences of the model families are obscured compared to aligning with respect to only the inspiral (i.e., with a frequency cut-off). 
Although this serves as clear evidence that the alignment strategy can overshadow intrinsic model differences, there is currently no direct means to distinguish between alignment residuals and intrinsic waveform dissimilarities. Hence, at this point, we simply emphasize the potential shortcomings of the alignment procedure used throughout this work, which computes~\eqref{eq:M-def} for $h_{2,2}$ over the full available time  (frequency) domain in concordance with the majority of the literature.

Lastly, Figures~\ref{fig:alignI}-\ref{fig:alignIV} clearly illustrate that the alignment procedure for precessing events exhibits significantly poorer performance compared to non-precessing events. 
Indeed, the alignment of precessing BBH mergers remains a subject of ongoing investigations within the GW community. The primary challenge arises from the need to identify a specific instance in time when the spins of both black holes point in the same direction for both the reference model and the one to be aligned to it.
However, due to the model-dependent length of the generated waveform, this point in time with congruent spin orientations might not even exist for any given event. 
Disregarding these instances in parameter space, precessing events can, in principle, be aligned using the mismatch function as outlined in equation~\eqref{equ:mismatch}, involving adjustments for $\phi_\text{ref}$ as well as four angles (two for each spin). 
Admittedly, it is not guaranteed that this procedure results in the correct congruent spin configuration. Instead, a physically similar orbital configuration sufficiently close with respect to the distance measure induced by~\eqref{equ:mismatch} might be chosen accidentally.

Despite the mentioned caveats, in this work, we employ the above procedure, computing the mismatch~\eqref{equ:mismatch} with respect to $\phi_\text{ref}$ as well as four spin angles for all precessing events. Unsurprisingly, the mismatch, on average, turns out to be much larger than for the non-precessing case. Overall, we found the outcome of the alignment procedure for precessing events to be poorly reliable.
As a consequence, in section~\ref{sec:analysis}, we will refrain from utilizing the alignment for precessing events. That is, we will not compare alignment dependent physical quantities in different models for precessing events but instead focus on other aspects of these events that are not (as) sensitive to the alignment process.


\section{Events under consideration}
\label{sec:events}
Given the substantial number of events considered in this study, it is instructive to discuss the covered regions in parameter space. This allows us to establish meaningful connections between our findings and phenomenological insights, especially in cases where one or more approximants exhibit large deviations with respect to the reference model. Additionally, by carefully spreading the considered events over a large parameter space, we avoid introducing selection biases. 

For each case, that is precessing and non-precessing BBH mergers, we discuss the considered events for which NR counterparts are available and simulate additional ones to diversify the parameter space under investigation. 

\subsection{Non-precessing events}
\begin{figure}
	\centering
	\includegraphics[width=0.9\linewidth]{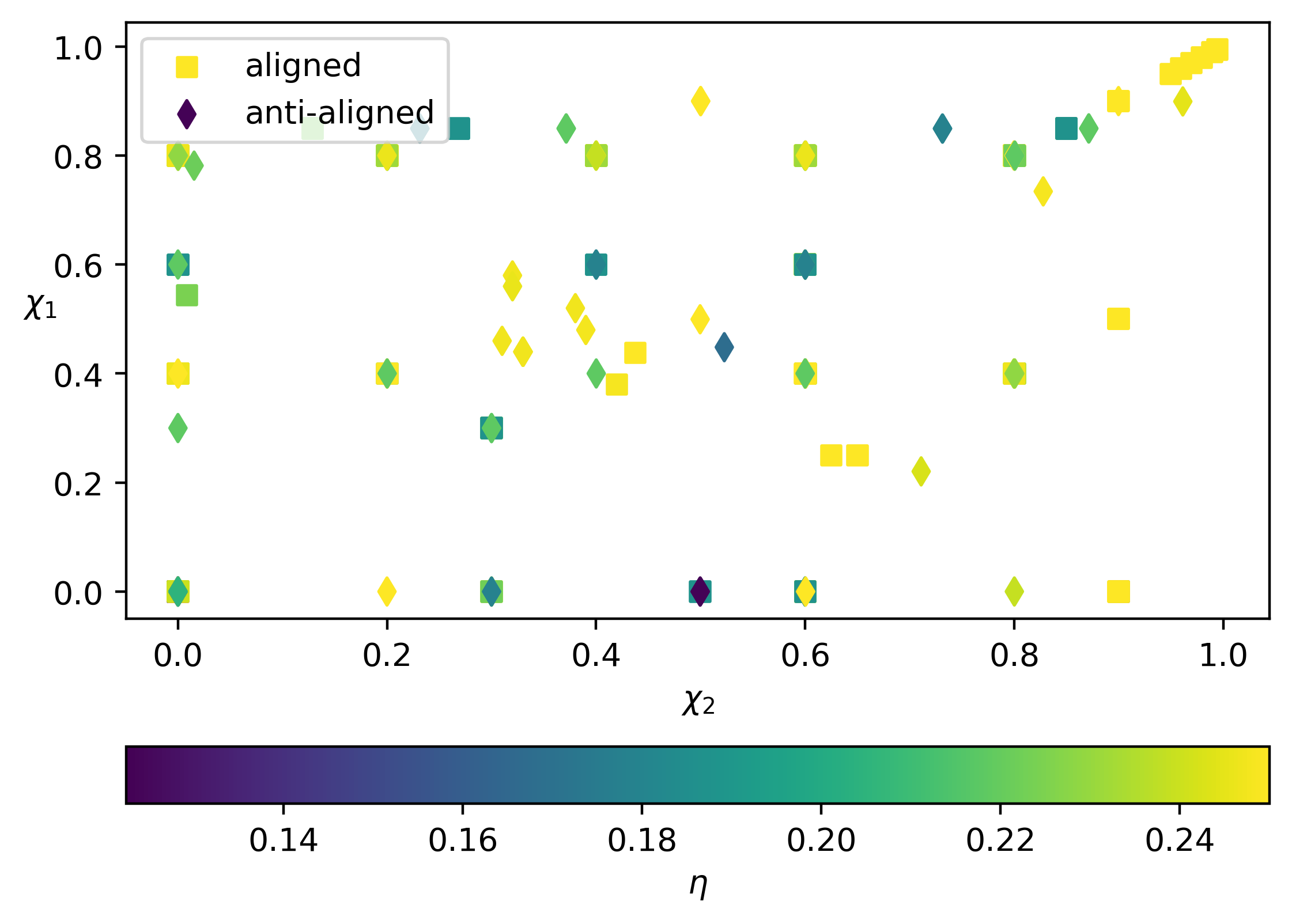}
	\caption{Parameter space for the non-precessing \SXS{} data.}
	\label{fig:params_I}
\end{figure}

The majority of our analysis resides in examining non-precessing events within the \SXS{} catalogue. In Figure~\ref{fig:params_I} we illustrate the distribution across parameter space, distinguishing between aligned and anti-aligned spin configurations and employing color-coding based on the mass ratio $\eta \coloneqq \frac{q}{(1+q)^2}$ with $q \coloneqq M_{i^+} / M_{i^\circ}$, corresponding to the ratio of final and initial mass. 
While a total of $175$ events with non-negligible kick, i.e., $v > 20 \text{km}/\text{s}$, is selected for the non-precessing case, the parameter space displayed in Figure~\ref{fig:params_I} seems to be only sparsely covered. This is due to many events depicted in Figure~\ref{fig:params_I} overlapping in the sense that either only their mass ratio or the alignment of spins changes. 

\begin{figure}
	\centering
	\includegraphics[width=0.9\linewidth]{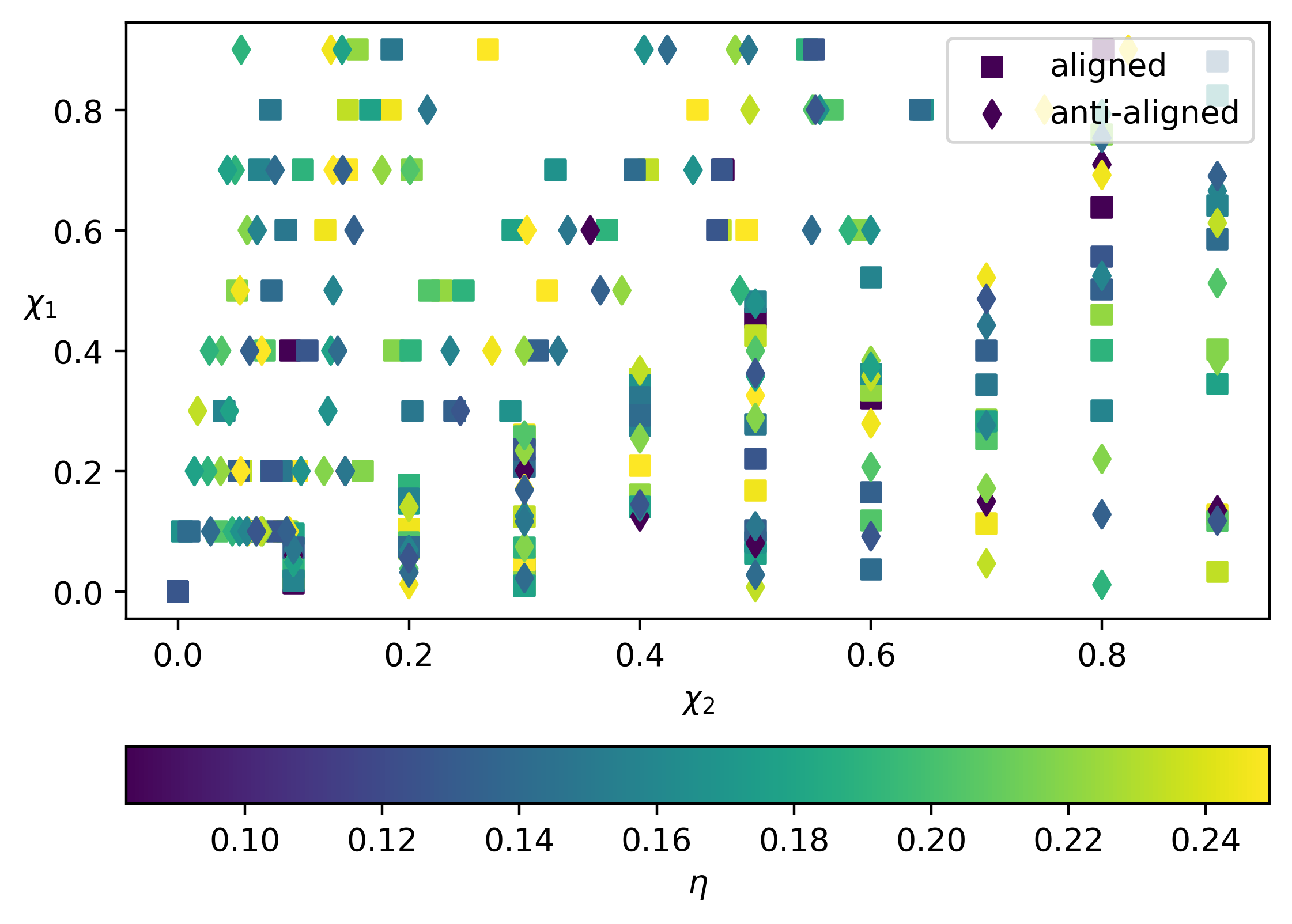}
	\caption{Parameter space for the non-precessing events without NR counterpart.}
	\label{fig:params_non_prec}
\end{figure}

From Figure~\ref{fig:params_I} it is evident that regions characterized by lower spins $\chi_1$ and $\chi_2$ are insufficiently represented within this excerpt of cataloged data. To address sparsely populated areas within the parameter space, we introduce additional simulated events without a \SXS{} counterpart. These events feature aligned as well as anti-aligned spins and span across an extensive range of mass ratios $\eta$, as depicted in Figure~\ref{fig:params_non_prec}. This augmented set adds an additional $220$ instances to our set of non-precessing events. 

Together, catalogued and non-catalogued non-precessing events form a well-distributed set, which is suitable for an unbiased systematic investigation.
Despite all efforts of accurately modeling GW waveforms, for some instances of non-precessing events, certain models may deviate strongly from the reference model and the other approximants in terms of remnant velocity or gravitational memory (again anticipating results of section~\ref{sec:analysis}). A closer investigation of such instances reveals that in these cases one of the modes of the deviating model (e.g. the $h_{2,1}$ mode) significantly differs from the corresponding mode of the remaining approximants and the reference model. This behavior predominantly occurs when one of the normalized spin components $\chi_{1,2}$ is extremal, i.e., either close to $0$ or $1$. Examples of these types of events include for instance \textit{SXS:BBH:0222}, \textit{SXS:BBH:0223}, and \textit{SXS:BBH:0251}, for which \EOB{} results in a kick velocity much larger or smaller than for the remaining waveform models. In order to not affect our statistical considerations, such events are excluded from our investigation.

\subsection{Precessing events}
\begin{figure}
	\centering
	\includegraphics[width=0.9\linewidth]{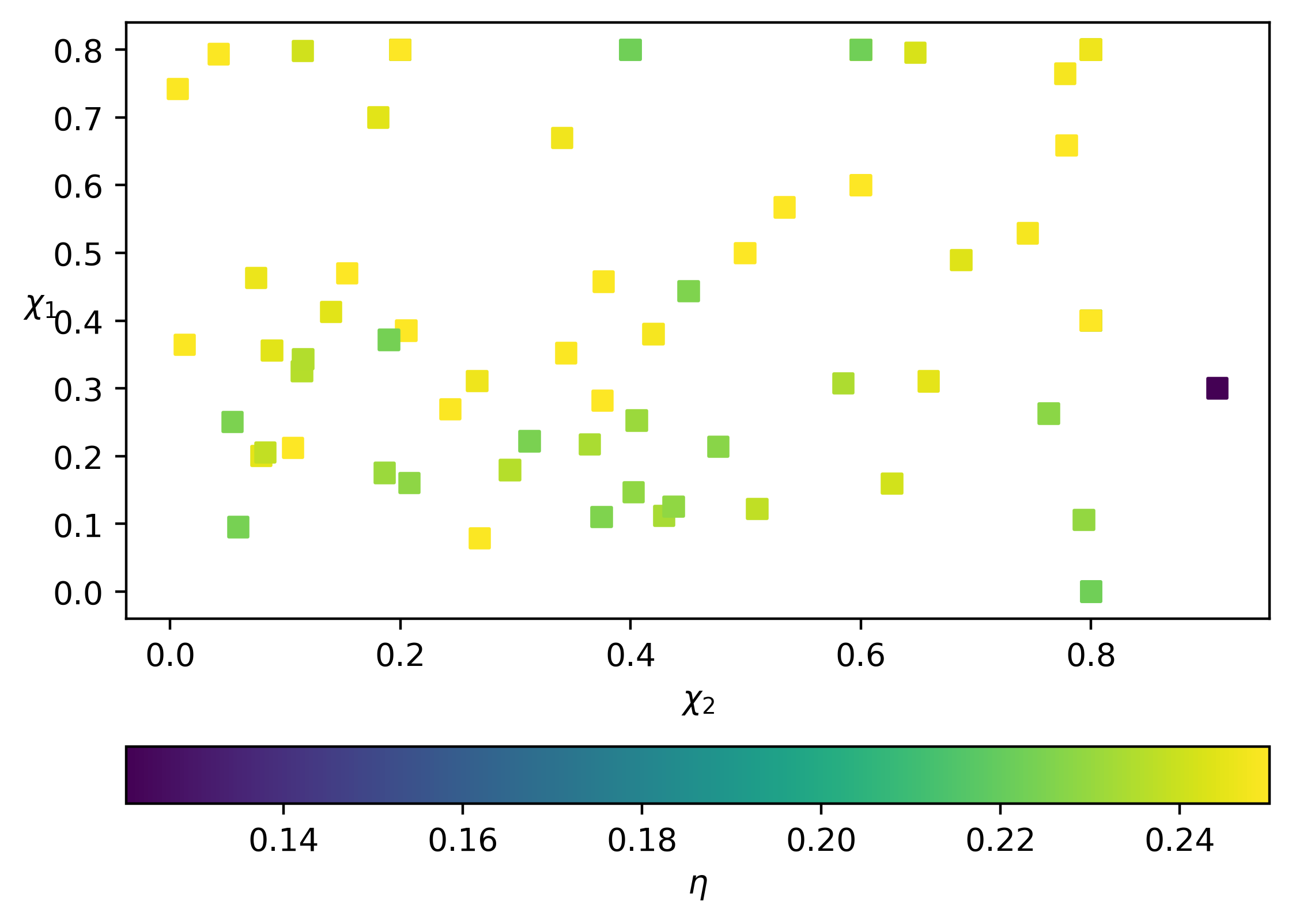}
	\caption{Parameter space for the precessing \SXS{} data.}
	\label{fig:params_prec_NR}
\end{figure}

\begin{figure}
	\centering
	\includegraphics[width=0.9\linewidth]{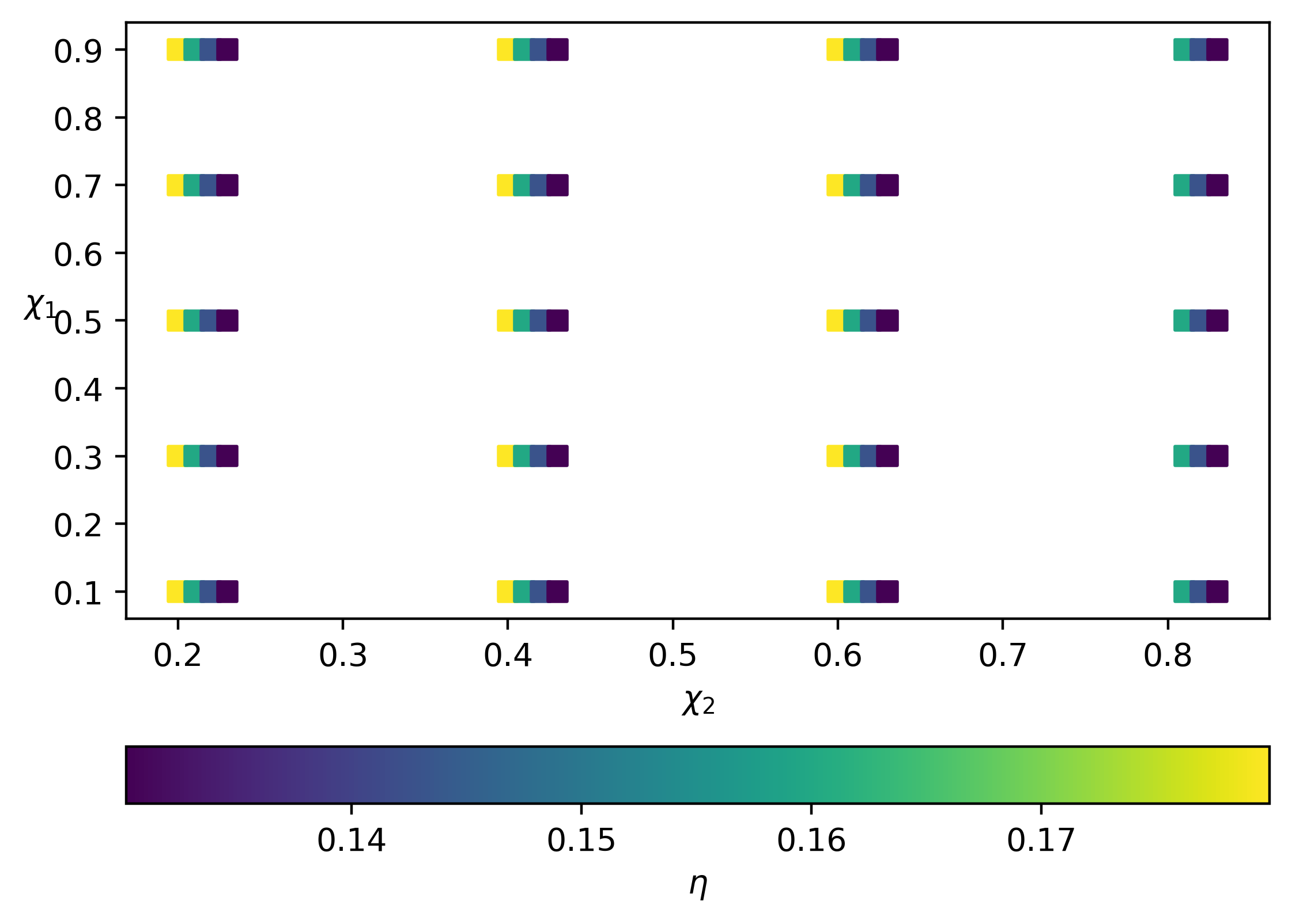}
	\caption{Parameter space for the precessing events without NR counterpart. For every bundle of data points, the spin configuration is identical. For illustrative reasons, the plot shows a slight misplacement.}
	\label{fig:params_prec}
\end{figure}

For the analysis of precessing BBH, we make use of $130$ events with \SXS{} counterparts, as displayed in Figure~\ref{fig:params_prec_NR}. In terms of the two spin vectors, the events are homogeneously distributed over the available parameter space. Notably, the catalogued events do not extend to lower mass ratios except for the event \textit{SXS:BBH:0165}, which involves two black holes of masses $51.4 \,\Msol$ and $8.6\, \Msol$. As this lack of low mass ratio binaries ($\eta<0.2$) potentially demonstrates a parameter bias, we supplement these precessing \SXS{} events with a set of $75$ additional instances. The added events are displayed in Figure~\ref{fig:params_prec}. The difference in number and randomness of the added events in the precessing and non-precessing case is a direct result of the computational complexity of aligning precessing events. For these events, we optimize the mismatch with respect to $5$ variables instead of only $\phi_\text{ref}$, resulting in a significantly larger data generation time. Due to time constraints, the added events in the precessing case are less numerous and more grid-like distributed. 


\section{Energy-momentum balance laws}
\label{sec:theory}
To assess a waveform model's performance against a chosen reference model, an adequate measure of comparison has to be selected. We choose the remnant's kick velocity and the gravitational wave memory, which are both physical observables that can be calculated solely from the strain of a GW.
In what follows we derive explicit expressions for the kick velocity and the memory from constraint equations which are known as \textit{energy-momentum balance laws}~\cite{Ashtekar:2019viz}. These laws are exact mathematical results derived from full, non-linear GR when applied to the coalescence of compact binaries. For an exhaustive review and derivation of the balance laws, we refer to~\cite{Ashtekar:2019viz, DAmbrosio:2022clk}. For a proof of concept that the balance laws can be used as diagnostic tools for assessing waveforms we refer to~\cite{Heisenberg:2023mdz}, which uses a simple analytical waveform model, and to~\cite{Khera:2020mcz}. Here, we will simply state the main result, explain its physical content, and perform a decomposition into spherical harmonics in order to derive explicit expressions for kick velocity and memory. These expressions then serve as the basis of our comparative analysis.

\subsection{The balance laws}
Waveform models describe GWs emitted from isolated systems composed of two compact objects which orbit each other and coalesce. The coalescence is caused by a loss of orbital energy, due to the emission of GWs. When the two compact objects finally merge, the remnant can be subjected to a ``kick''. This is a consequence of the fact that GWs are emitted anisotropically and because they do not only carry energy, but also momentum. Based on this simple qualitative picture of the physics of compact binary coalescence, one would expect that there is a mathematical law which describes the energy loss of the system and balances it against the energy carried away by GWs. One may furthermore expect that this law establishes a relation between \textit{(i)} the initial and final mass of the system, \textit{(ii)} the kick velocity, and \textit{(iii)} the strains $h_+$, $h_\times$ of the GW. Such a mathematical law does indeed exist, and it can be derived from full, non-linear GR when applied to compact binary coalescence~\cite{Ashtekar:2019viz}. As a matter of fact, there is not only one law, but infinitely many. Namely, one per point on the $2$-sphere, i.e., one so-called energy-momentum balance law per choice of $(\theta, \phi)$. These laws can be brought into the form
\begin{widetext}
\begin{align}\label{equ:fullDimBL}
    c^2 \left(\frac{M_\text{remnant}}{\gamma^3\left(1-\frac{\vec{v}}{c}\cdot\hat{x}\right)^3} - M_\text{binary}\right) = -\frac14 \frac{D^2_L\, c^3}{G}\int_{-\infty}^{\infty} \left(\dot{h}^2_+ + \dot{h}^2_\times\right)\dd t + \frac12 \frac{D_L\, c^4}{G} \text{Re}\left[\eth^2\left(h_+ - i\, h_\times\right)\right]\bigg|_{t=-\infty}^{t=+\infty}\,,
\end{align}
\end{widetext}
where $\vec{v}$ stands for the kick velocity, $\gamma \equiv \gamma(v)$ is the usual Lorentz factor from Special Relativity, $\hat{x} = (\sin\theta \cos\phi, \sin\theta \sin\phi,\cos\theta)$ is the unit radial vector in spherical coordinates, $D_L$ is the luminosity distance between source and observer, and $\eth$ is a differential operator called ``eth'' whose precise definition can be found in~\cite{Ashtekar:2019viz,DAmbrosio:2022clk}. 

By recalling our qualitative description of the physics of compact binary coalescence, we can give a tentative interpretation to the balance laws. Since $E = M\, c^2$, according to Einstein, we can interpret the left-hand side of~\eqref{equ:fullDimBL} as the difference of two energies measured at two different instants of time. Namely, the energy of the remnant, which corresponds to the energy of the system after the merger, minus the energy of the binary system, which is the energy measured well before the merger. The denominator containing the Lorentz factor and the kick velocity can be understood as a correction due to the fact that the energy of the binary system is measured in its instantaneous rest frame, while the remnant is, in general, moving with respect to that frame with a velocity $\vec{v}$. As we will see below, when we decompose~\eqref{equ:fullDimBL}, for the $\ell=0$ mode this factor reduces precisely to the special relativistic formula for the kinetic energy of a body moving relative to an inertial frame, $E=\gamma(v) M c^2$. For a derivation of this factor and technical details we refer the reader to~\cite{Ashtekar:2019viz}. Since the binary system is losing energy due to the emission of GWs, it is natural to expect that the difference of energies is negative. Indeed, on the right-hand side of~\eqref{equ:fullDimBL} we find an integral over a manifestly positive quantity, $\left(\dot{h}^2_+ + \dot{h}^2_\times\right)$, with a minus sign in front. This integral computes the energy carried by GWs, and it is a well-known result of linearized GR. However, we stress again that these balance laws hold well beyond the linear regime and, in fact, no linearization is ever used. Finally, the second term on the right-hand side represents the GW memory. Because the left side of~\eqref{equ:fullDimBL} depends on $\hat{x}$, which is a function of $(\theta, \phi)$, and because the strains $h_+$, $h_\times$ are not only functions of time but also functions of the angular variables $(\theta, \phi)$, we find that there is indeed one balance law per choice of $(\theta, \phi)$. 

The strength and diagnostic capabilities of the balance laws derive from the fact that they are a precise mathematical result which is valid in full GR. The quantities which appear in their formulation, i.e., initial and final mass of the binary system, kick velocity, and the GW strains, are all quantities which either enter as parameters or are provided in an approximate fashion by waveform models. Thus, the exact balance laws can be used to test the accuracy of waveform models.

The balance laws are based on the pioneering work of Bondi, Metzner, Sachs, and van der Burg, who laid the foundations for describing GWs beyond the linear approximation~\cite{bondiGravitationalWavesGeneral1962, sachsGravitationalWavesGeneral1962}. Their work was subsequently extended and formalized by Newman, Penrose, Geroch, Ashtekar, and others, who introduced precise mathematical definitions for ``asymptotically flat'' and ``asymptotically Minkowski'' spacetimes~\cite{penroseAsymptoticPropertiesFields1963, gerochAsymptoticStructureSpaceTime1977, ashtekarGeometryPhysicsNull2015}, which, qualitatively speaking, capture the idea of having an isolated system contained within a finite spacetime region. An other key element in the derivation of the balance laws are the symmetries of asymptotic Minkowski spacetimes. These are described by the so-called Bondi-Metzner-Sachs (BMS) group~\cite{bondiGravitationalWavesGeneral1962}. By definition, this group leaves the asymptotic structure on null infinity $\scrip$ (read: scri plus) invariant. A variant of Noether's theorem can then be applied, which leads to a description of the (non-)conservation of Noether charges: The energy-momentum balance laws~\cite{Ashtekar:1981, waldGeneralDefinitionConserved2000}. The balance laws~\eqref{equ:fullDimBL} shown here are an adaptation to the special case of compact binary coalescence. We refer to~\cite{DAmbrosio:2022clk} for a didactic introduction and recent review of the fascinating mathematical subjects (asymptotic Minkowski spacetimes, the BMS group, balance laws), which here we only brushed over.

In what follows, we will need the mode-decomposition of equation~\eqref{equ:fullDimBL} with respect to spherical harmonics $Y_{\ell m}(\theta, \phi)$. In order to compactify the notation, we set\footnote{This is standard notation in the balance law literature and it hints at its origin: The balance laws formally hold on null infinity, $\scrip$. If one takes the limit into the distant past, i.e., well before the binaries merge, one reaches spacelike infinity, $i^\circ$. On the other hand, if one moves along $\scrip$ into the distant future, well after the binaries have merged, one reaches timelike infinity, denoted by $i^+$.}
\begin{align}
    M_\text{binary} &\coloneqq M_{i^\circ} & \text{and} && M_\text{remnant} &\coloneqq M_{i^+}\,,
\end{align}
and we also introduce the asymptotic shear
\begin{align}
    h(t, \theta, \phi) \coloneqq \frac12 \left(h_+ + i\, h_\times\right)(t, \theta, \phi)\,,
\end{align}
where $i$ is the imaginary unit. This last definition in particular allows us to re-write the memory term compactly as $\eth^2\Delta\bar{h}$, with $\bar{h}$ denoting the complex conjugate of $h$ and where
\begin{align}
    \Delta \bar{h} \coloneqq \int_{-\infty}^{+\infty} \dot{\bar{h}}\, \dd t\,.
\end{align}
Re-writing the memory term in this fashion is possible thanks to the fact that $\text{Im}(\eth^2 \bar{h})$ vanishes at $t=\pm \infty$ (cf.~\cite{Ashtekar:2019viz}).

Using these definitions and expanding both sides of equation~\eqref{equ:fullDimBL} into spherical harmonics, we obtain a tower of constraints, namely one for each pair of $(\ell, m)$ with $|m|\leq \ell$. Concretely, this decomposition reads
\begin{align}\label{equ:BLinModes}
    \left(M_{i^\circ} - \frac{M_{i^+}}{\gamma^3\left(1-\frac{\vec{v}}{c}\cdot \hat{x}\right)^3}\right)_{\ell,m} &= \frac{D^2_L c}{G}\int_{-\infty}^{+\infty}\left(|\dot{h}|^2\right)_{\ell,m}\dd t\notag\\
    &\phantom{=} - \frac{D_L c^2}{G} C_\ell \Delta \bar{h}_{\ell, m}\,,
\end{align}
where the $C_\ell$ coefficients are defined as
\begin{align}
    C_\ell \coloneqq \sqrt{(\ell-1)\ell(\ell+1)(\ell+2)} \,.
\end{align}
To arrive at this expression, we used well-known properties of spherical harmonics and the explicit expression of $\eth^2\bar{h}$, which can for instance be found in~\cite{DAmbrosio:2022clk}. Note that $C_\ell$ tells us that not each term in the balance laws~\eqref{equ:fullDimBL} contributes to each mode. In fact, $C_\ell$ vanishes for $\ell = 0$ and $\ell=1$. Thus, the memory term only contributes for $\ell \geq 2$. Treating the $\ell < 2$ and $\ell\geq 2$ modes of the balance laws separately allows us to devise a method for computing the remnant mass, in case it is not provided by the waveform model, the kick velocity, and the memory solely based on the GW strain.

For carrying out the mode-decomposition of the left hand side of equation~\eqref{equ:BLinModes}, it is convenient to first align the $z$-axis of the rest frame in which $M_{i^\circ}$ is measured with $\vec{v}$ using a rotation in the plane spanned by $\vec{v}$ and $\hat{z}$. The first few $(\ell,m)$ coefficients then read
\begin{align}
    (\ell = 0, m=0)&: & &2 \sqrt{\pi } (\gamma  M_{i^+} - M_{i^\circ}) ,\notag\\
    (\ell = 1, m=0)&: & &2 \sqrt{3 \pi } \gamma  M_{i^+} v,\notag \\
    (\ell = 2, m=0)&: & &\frac{\sqrt{5 \pi } \gamma M_{i^+} \left(5 v^3+3 \gamma^{-4} \tanh ^{-1}(v) -3 v\right)}{v^3} \,.\notag
\end{align}
Observe that the first term is, up to a factor of $c^2$, simply the difference between the rest energy $M_{i^\circ}$ of the binary and the kinetic energy $\gamma\, M_{i^+}$ of the remnant, while the second term is the ($z$-component) of the momentum of the remnant. The third term does not admit a simple physical interpretation, but we emphasize that the inverse of the hyperbolic tangent of $v$ caused numerical instabilities in the code we used for our analysis. We therefore Taylor expanded $\tanh^{-1}(v)$ up to sixth order in $v/c$. 
To obtain the corresponding modes in the original frame, we transform the above modes using Wigner $D$-matrices $D^\ell_{m,m'}$ for the inverse rotation used to align $\vec{v}$ and $\hat{z}$.

The decomposition of the GW energy integral in~\eqref{equ:BLinModes} can easily be carried out. In fact, the integrand can be written as
\begin{equation}\label{equ:strain_decompose}
    |\dot h|^2 = \sum_{\ell, m} \alpha_{\ell m} Y_{\ell m}(\theta, \phi)\,.
\end{equation}
A straightforward computation reveals that the $\alpha_{\ell m}$ coefficients are given by (see also~\cite{Khera:2020mcz})
\begin{widetext}
\onecolumngrid
    \begin{align}\label{equ:alphas}
        \alpha_{\ell m} = \sum_{\ell_1=2}^\infty \sum_{\ell_2=2}^\infty \sum_{|m_1|\leq \ell_1}\sum_{|m_2|\leq \ell_2} (-1)^{m_2+m} \dot h_{\ell_1 m_1} \dot{\bar{h}}_{\ell_2m_2}\sqrt{\frac{(2\ell_1+1)(2\ell_2+1)(2\ell+1)}{4\pi}}\begin{pmatrix}
\ell_1 & \ell_2 & \ell\\
m_1 & -m_2 & -m
\end{pmatrix}\begin{pmatrix}
\ell_1 & \ell_2 & \ell\\
2 & -2 & 0
\end{pmatrix}.
    \end{align}
\end{widetext}

The $\begin{psmallmatrix} \ell_1 & \ell_2 & \ell\\ m_1 & m_2 & m\end{psmallmatrix}$ denotes the Wigner-$3j$ symbol, which effectively determines which strain modes $\dot h_{\ell_1 ,m_1} \dot{\bar{h}}_{\ell_2,m_2}$ couple to each other in~\eqref{equ:alphas}.

\subsection{Remnant mass and velocity}
For astrophysically realistic kick velocities one can assume $\gamma\approx 1$. This approximation reduces the $\ell = 0$ mode of the balance laws to an energy conservation equation of the form
\begin{align}\label{equ:Energy}
    c^2 (M_{i^\circ}-M_{i^+}) = \frac{D_L^2 c^3}{16\pi G}\int_{-\infty}^{\infty }\dd t \oint \dd \Omega \,|\dot h|^2\,,
\end{align}
where $\oint\dd\Omega$ is an integral over the unit $2$-sphere. In agreement with our intuition, the mass loss of the system is accounted for by the energy radiated away by GWs. For waveform models which provide the mass of the remnant, this formula can be used as a consistency check. On the other hand, when a waveform model does not provide $M_{i^+}$, the above formula can be used to determine the remnant's mass using only the total mass of the binary system, the luminosity distance, and the GW strain as input. Upon using the expansion coefficient~\eqref{equ:alphas} for $\ell=m=0$, the energy conservation equation can also be expressed as
\begin{align}
    c^2 (M_{i^\circ}-M_{i^+}) &=\frac{D_L^2 c^3}{8\sqrt{\pi} G}\int_{-\infty}^{\infty}\dd t\,\alpha_{0,0}.
\end{align}
The $\ell = 1$ mode of the balance laws encode conservation of linear momentum, as one might have expected. This allows us to solve the momentum equations for the components of the kick velocity, which can be expressed as
\begin{equation}\label{equ:kick}
    v_i = \frac{D_L^2 c^2}{16 \pi G M_{i^+}} \int_{-\infty}^\infty \dd t \oint \dd \Omega \, \hat{x}_i\, |\dot h|^2,
\end{equation}
in the original reference frame (i.e., the frame where $\vec{v}$ and $\hat{z}$ are not aligned). Here, $\hat{x}_i$ stand for the $i$-th component of the radial unit vector $\hat{x} = (\sin\theta\cos\phi,\sin\theta\sin\phi,\cos\theta)$. The kick velocity components can of course also be expressed in terms of the $\alpha_{\ell m}$ coefficients for $\ell=1$ and $m=-1,0,+1$. One obtains
\begin{align}
    v_1 &= \frac{D_L^2 c^2}{16 \pi G M_{i^+}} \sqrt{\frac{2\pi}{3}} \int_{-\infty}^\infty \dd t \, (\alpha_{1,-1} - \alpha_{1,1})\label{equ:v1} \\
    v_2 &= \frac{-iD_L^2 c^2}{16 \pi G M_{i^+}} \sqrt{\frac{2\pi}{3}} \int_{-\infty}^\infty \dd t \, (\alpha_{1,-1} + \alpha_{1,1}) \\
    v_3 &= \frac{D_L^2 c^2}{8 \pi G M_{i^+}} \sqrt{\frac{\pi}{3}} \int_{-\infty}^\infty \dd t \, \alpha_{1,0} \,.\label{equ:v3} 
\end{align}
Equations~\eqref{equ:v1}-\eqref{equ:v3} determine the kick velocity using luminosity distance, remnant mass, and strain as input. If $M_{i^+}$ is not provided by a waveform model, it can be determined via the energy conservation equation~\eqref{equ:Energy}. 

\subsection{Gravitational memory}
As we have seen above, modes with $\ell\geq 2$ contain contributions from the GW memory. The memory term can be decomposed into two contributions,
\begin{equation}
\label{eq:mem_total_def}
 \Delta \bar{h}_{\ell,m} = \Delta \bar{h}^\text{lin}_{\ell,m} + \Delta \bar{h}^\text{non-lin}_{\ell,m}\,.
\end{equation}
Following standard conventions in the literature, we define the first contribution as
\begin{equation}
\label{eq:lin_mem_def}
    \Delta \bar{h}^\text{lin}_{\ell,m} \coloneqq \frac{G}{C_\ell D_L c^2 }\left(\frac{M_{i^+}}{\gamma^3(1 - \frac{\vec {v}}{c}\cdot \hat{x})^3} - M_{i^\circ}\right)_{\ell,m}
\end{equation}
and call it the \textit{linear memory}~\cite{zeldovichRadiationGravitationalWaves1974, braginskyGravitationalwaveBurstsMemory1987}. The second term, defined as 
\begin{align}
\label{eq:nonlin_mem_def}
\Delta \bar{h}^\text{non-lin}_{\ell,m} \coloneqq  \frac{D_L}{4C_\ell c}\int_{-\infty}^\infty \dd t \, \alpha_{\ell m}
\end{align}
is known as the \textit{non-linear memory}~\cite{christodoulouNonlinearNatureGravitation1991a, thorneGravitationalwaveBurstsMemory1992}.
Other names for these two terms are \textit{ordinary memory} for the former and \textit{null memory} for the latter~\cite{Mitman:2020bjf}. 
More importantly, we interpret the $\ell\geq 2$ modes of the balance laws~\eqref{equ:BLinModes} as constraints on the memory terms. The same constraints have been used in previous studies to either add memory to waveform models or correct waveform models which did not accurately incorporate the memory effect~\cite{Khera:2020mcz, Mitman:2020bjf}. In this work, we instead use the linear and non-linear memory inferred from~\eqref{eq:lin_mem_def} and~\eqref{eq:nonlin_mem_def}, respectively, as a means of comparison between different waveform models. 

To summarize this section, we decomposed the balance laws~\eqref{equ:fullDimBL} into spherical harmonic modes to obtain analytical equations for the remnant mass, the kick velocity, and the GW memory. The remnant mass $M_{i^+}$ is inferred from the $\ell=0$ mode of the balance laws, while the kick velocity is deduced from the conservation of linear momentum implied by the $\ell=1$ mode. The $\ell\geq 2$ modes are used to infer linear and non-linear GW memory. Equations~\eqref{equ:Energy}, \eqref{equ:kick}, \eqref{eq:lin_mem_def}, and \eqref{eq:nonlin_mem_def} are implemented numerically and applied to all the models and events considered in this work. The resulting waveform assessment is the subject of the next section.


\section{Waveform Assessment}
\label{sec:analysis}
Computing the remnant velocity following equations \eqref{equ:v1}-\eqref{equ:v3} and evaluating the gravitational memory, encoded in equations \eqref{eq:lin_mem_def} and \eqref{eq:nonlin_mem_def}, we assess the performance of the chosen waveform approximants across the parameter space specified in~\ref{sec:alignmentandparas}.  Specifically, we examine the impact of the subdominant strain modes on the kick velocity and gravitational wave memory. We restrict our analysis of alignment-sensitive quantities to non-precessing events only.
\begin{figure}
	\centering
 \includegraphics[width=0.99\linewidth]{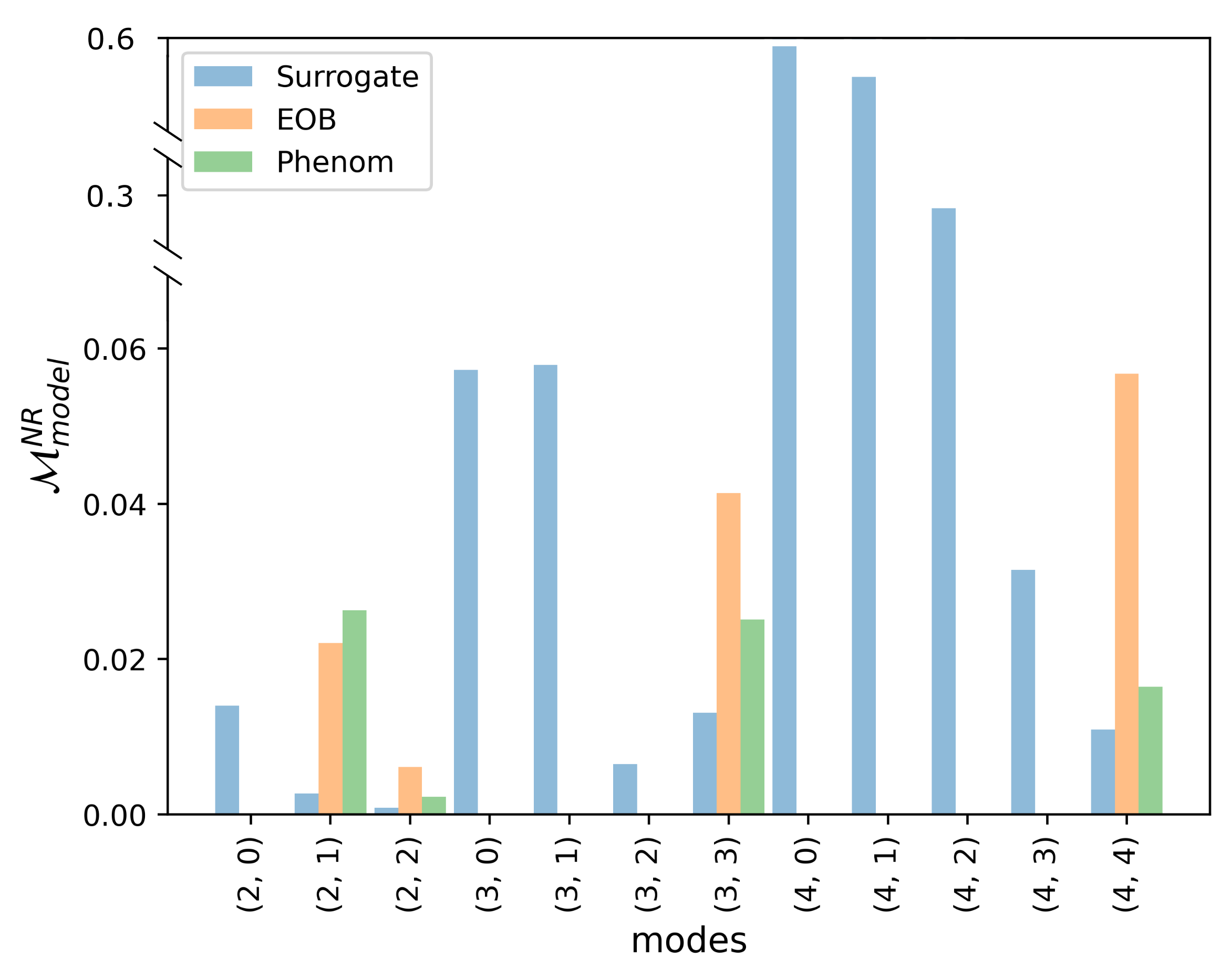}
	\caption{Mode mismatch with respect to NR, averaged over all non-precessing events.}
	\label{fig:mismatch_NR_np}
\end{figure}
\subsection{Mode Mismatch}
We start by considering the non-precessing events detailed in section~\ref{sec:events}, for which NR waveforms are accessible. 
A natural measure of comparability between waveforms (or individual modes) is the mismatch function~\eqref{equ:mismatch}. In the alignment procedure, the mismatch function $\mathcal{M}$ is minimized for the dominant $h_{2,2}$ mode. The remaining modes are not exposed to any direct mismatch minimization. Thus, a first intuition about potential shortcomings of waveform models can be gained based on the residual mismatch of the subdominant modes, and how the latter compares to the mismatch in the dominant $h_{2,2}$. In Figure~\ref{fig:mismatch_NR_np}, we present the mismatch calculated via~\eqref{equ:mismatch} for the mode content common to all waveform models\footnote{Note that we display only up to $\ell=4$ due to the restricted mode content of \Surr{}.}, averaged over all events under consideration. 
As expected, we observe a low mismatch for the dominant $h_{2,2}$ mode throughout approximants, with \Surr{} performing best in this metric. For subdominant $\ell$ modes, the mismatch becomes larger. Particularly for $\ell=4$ modes but also for $h_{2,0}$, $h_{3,2}$, $h_{3,1}$, $h_{3,0}$ and their complex conjugates, the mismatch is multiple orders of magnitude larger than for $h_{2,2}$. The higher mismatch between \Surr{} and NR in subdominant modes is attributed to \Surr{} prioritizing the modelling accuracy of the dominant $h_{2,2}$ in its interpolation procedure.

For multiple subdominant modes, \EOB{} and \Phen{} obtain trivial contributions. As this would result in a mismatch $\mathcal{M}=1$, we choose not to include those instances for \EOB{} and \Phen{} in Figure~\ref{fig:mismatch_NR_np}. The absence of these modes extends to precessing events as well, despite internal procedures modifying the mode content via, for instance, the twisting-up procedure in \Phen{}, as outlined in section~\ref{subsubsec:precession}. However, the methods by which precessing waveforms are obtained from non-precessing ones cannot compensate the lack of information stored in the subdominant modes of a model. Thus, for precessing events a similar pattern regarding the mode-by-mode mismatches as in Figure~\ref{fig:mismatch_NR_np} can be found. 

We conducted a similar analysis for non-catalogued events, i.e., events for which no NR waveform is available. In this case, we chose \Surr{} as the reference model and observe an analogous relationship between the mismatch of \EOB{} and \Phen{} versus \Surr{}. Specifically, for $\ell=2$, the mismatch with respect to \Surr{} for both \EOB{} and \Phen{} is low. However, for $h_{3,3}$ and $h_{4,4}$, \EOB{} waveforms exhibit visibly stronger dissimilarities with respect to \Surr{} than \Phen{} does.
\begin{figure}
   \includegraphics[width=0.9\columnwidth]{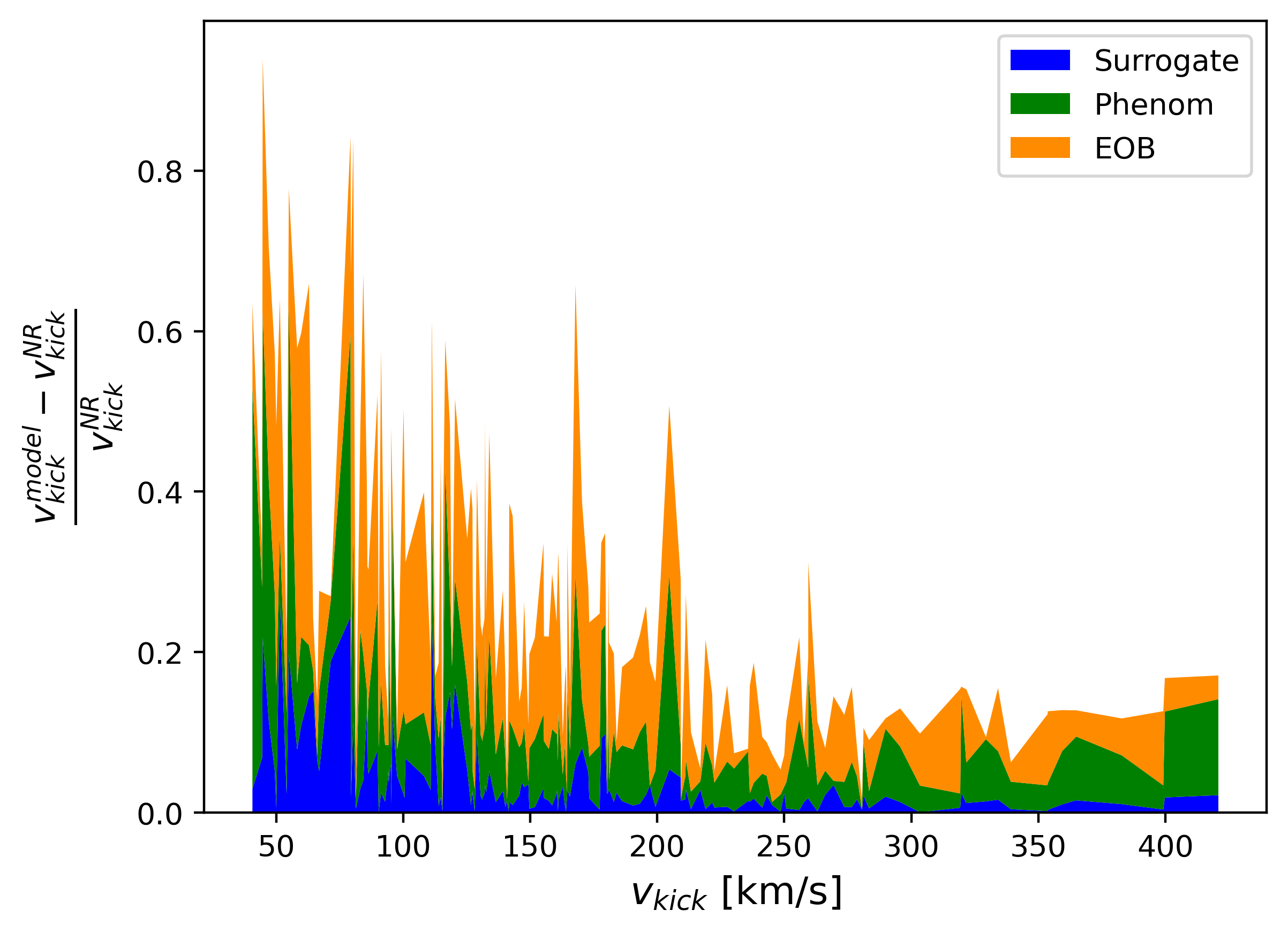}    
\includegraphics[width=0.8\columnwidth]{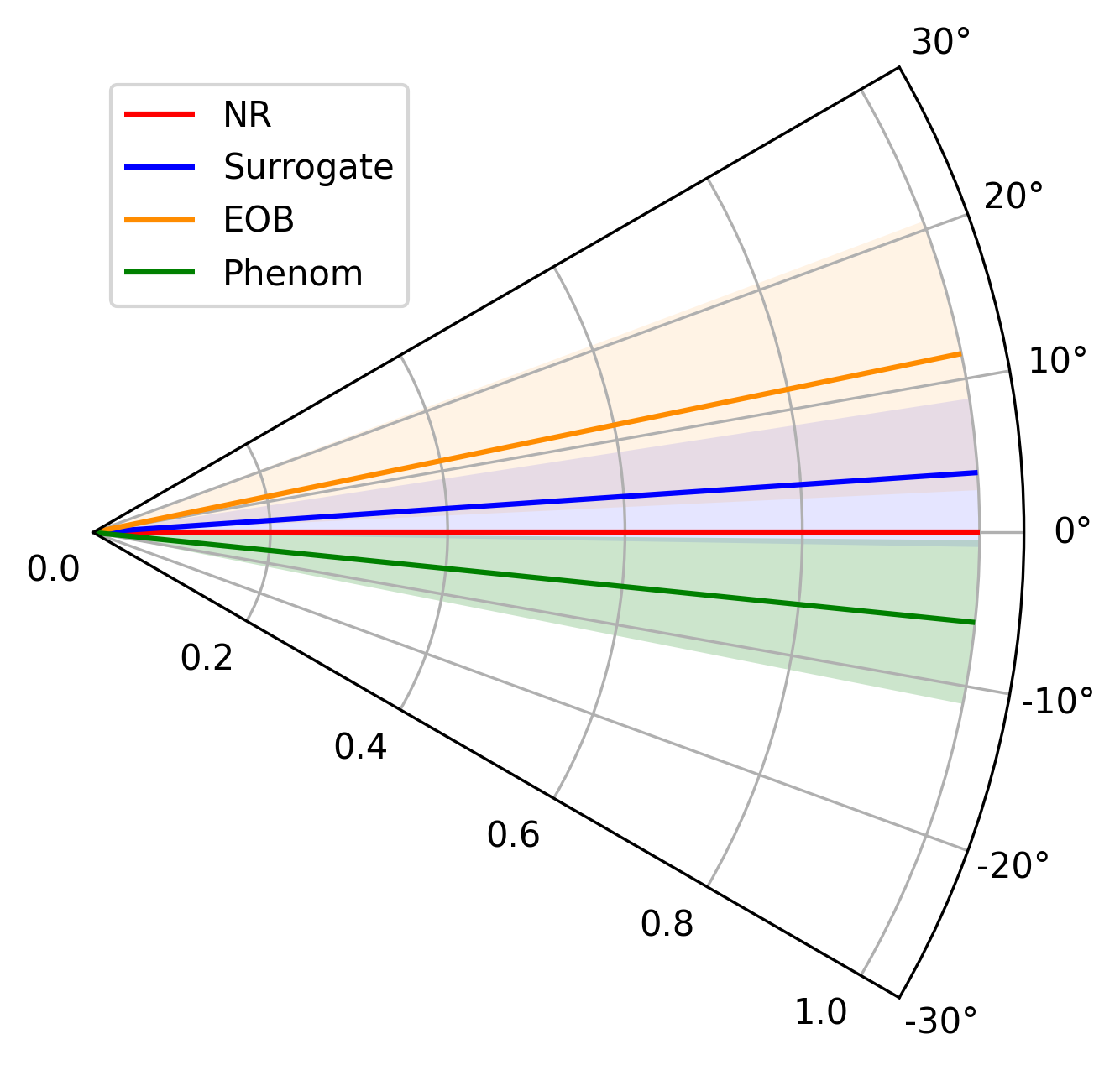}   
     \caption{Relative error for the kick magnitude (top) and absolute deviation angle (bottom) of the selected approximants with respect to NR for non-precessing events. The top stack plot shows the relative errors for each individual event labeled by its kick magnitude. The bottom plot shows the mean deviation over all tested events with respect to NR. A $2\sigma$-interval around the mean is colorized correspondingly.}
       \label{fig:error_kick_NR_np}
\end{figure}

In addition to intrinsic differences in the waveform modes, the mismatch might be influenced by what we term ``numerical artifacts'', as discussed in appendix~\ref{sec:numerical_artefacts}. These artifacts, particularly prominent in higher modes, require separate consideration. In the literature, some artifacts similar in nature to what is displayed in Figure~\ref{fig:artefact} have been addressed and mitigated, for instance, through the alignment of BMS charges~\cite{Mitman:2022kwt}. Here, we will not delve into a detailed discussion about their origin or impact but simply point out that even with a well-defined alignment procedure, the numerical nature of the analysis can potentially affect the mismatch results.

Despite the higher mismatch of certain modes, the full waveforms provide an adequate approximation to NR due to the dominance of the well-modeled $h_{2,2}$ mode for the majority of events. At the current level of precision, addressing the individual subdominant mode mismatch is thus only of secondary interest. However, in the perspective of future GW probes, the precision benchmark will increase significantly, potentially rendering higher modes more relevant~\cite{pitte_detectability_2023}. Hence, it is imperative for future GW waveform models to confront the evident challenges depicted in Figure~\ref{fig:mismatch_NR_np}, such as incomplete mode contents and decreasing precision of higher-order (subdominant) modes. In the subsequent discussion, we establish the groundwork to address these issues for individual waveform models by examining the feasibility of subdominant mode testing using the analytical tools presented in section~\ref{sec:theory}. 

\subsection{Testing Subdominant Modes}
In order to investigate the impact of subdominant modes, we analyze physical quantities of BBH mergers with regard to their sensitivity towards mode content. Concretely, we compute the relevant quantities using two sets of strain modes. The first set includes all available modes for each model, the second set does only include high-mismatch subdominant modes, e.g. we exclude modes belonging to the set\footnote{Note that modes that exceed the mode content of \Surr{} model, i.e., $\ell>4$ are not excluded for the remaining models (including NR).}
\begin{equation*}
H_{\text{sub}}=\{h_{2,0},h_{3,\pm 2},h_{3,\pm1},h_{3,0},h_{4,\pm 3},h_{4,\pm2}, h_{4,\pm1}, h_{4,0}\}.
\end{equation*}

\subsubsection{Remnant Velocity}
We first consider the recoil velocity of the remnant BH. Our attention thereby is limited to events with kick velocities varying between $20$ and $400$ ${\text{km}/}{\text{s}}$. Events with a negligible kick (i.e., $v<20$ ${\text{km}/}{\text{s}}$) are excluded from the analysis. We compare the magnitude and direction of the remnant velocities~\eqref{equ:kick} by computing each models' relative error with respect to NR. Computed using the full mode content, the relative error for the kick magnitude and the directional deviation of the remnant velocity vectors are displayed in Figure~\ref{fig:error_kick_NR_np}. In the top plot, we stack the relative errors and plot them against the corresponding kick magnitude for all events. The bottom plot displays the average of the absolute deviation in terms of kick direction with respect to NR. Negative angles for \Phen{} are assigned for better visibility in the plot.

From a quantitative point of view, Figure~\ref{fig:error_kick_NR_np} shows a good agreement between \Surr{} and NR, as expected. In particular, for larger velocities, the relative errors are small and the directional deviation is less than $5$ degrees on average. \EOB{} and \Phen{} waveforms cannot reproduce the reference velocity vector with comparable accuracy. They also display a similar trend of larger relative errors for lower remnant velocities. The strong fluctuations indicate a higher dependency on the individual event and its corresponding extrinsic parameters. For both the kick magnitude and direction, \EOB{} exhibits on average the largest errors throughout the tested parameter space. In total, these results agree with the findings of previous assessments based on kick velocities~\cite{Borchers:2021vyw}.

Computing similar statistics without the subdominant high-mismatch modes $H_{\text{sub}}$, we find that neither the kick magnitude nor its direction exhibit significant differences compared to the values computed using the full mode content. For NR and \Surr{}, the change in directional deviation and the relative magnitude error, on average, corresponds to only a few percent of the corresponding full-mode content values. For \EOB{} and \Phen{} we do not find statistically significant differences between excluding or including $H_{\text{sub}}$. The same applies to non-catalogued non-precessing events.

Our results indicate that quantities related to the remnant velocity of the final BH are not sensitive to the subdominant mode content of the strain produced by a GW waveform model. This observation extends to the computation of the final mass of the remnant black hole (cf. equation~\eqref{equ:Energy}). The insensitivity of the remnant's mass and kick velocity with respect to the set of modes $H_{\text{sub}}$ can be readily explained by the analytical expressions~\eqref{equ:alphas} for the $\alpha$-coefficients and equations~\eqref{equ:v1}-\eqref{equ:v3}:
\begin{figure}
	\centering
	\includegraphics[width=0.95\linewidth]{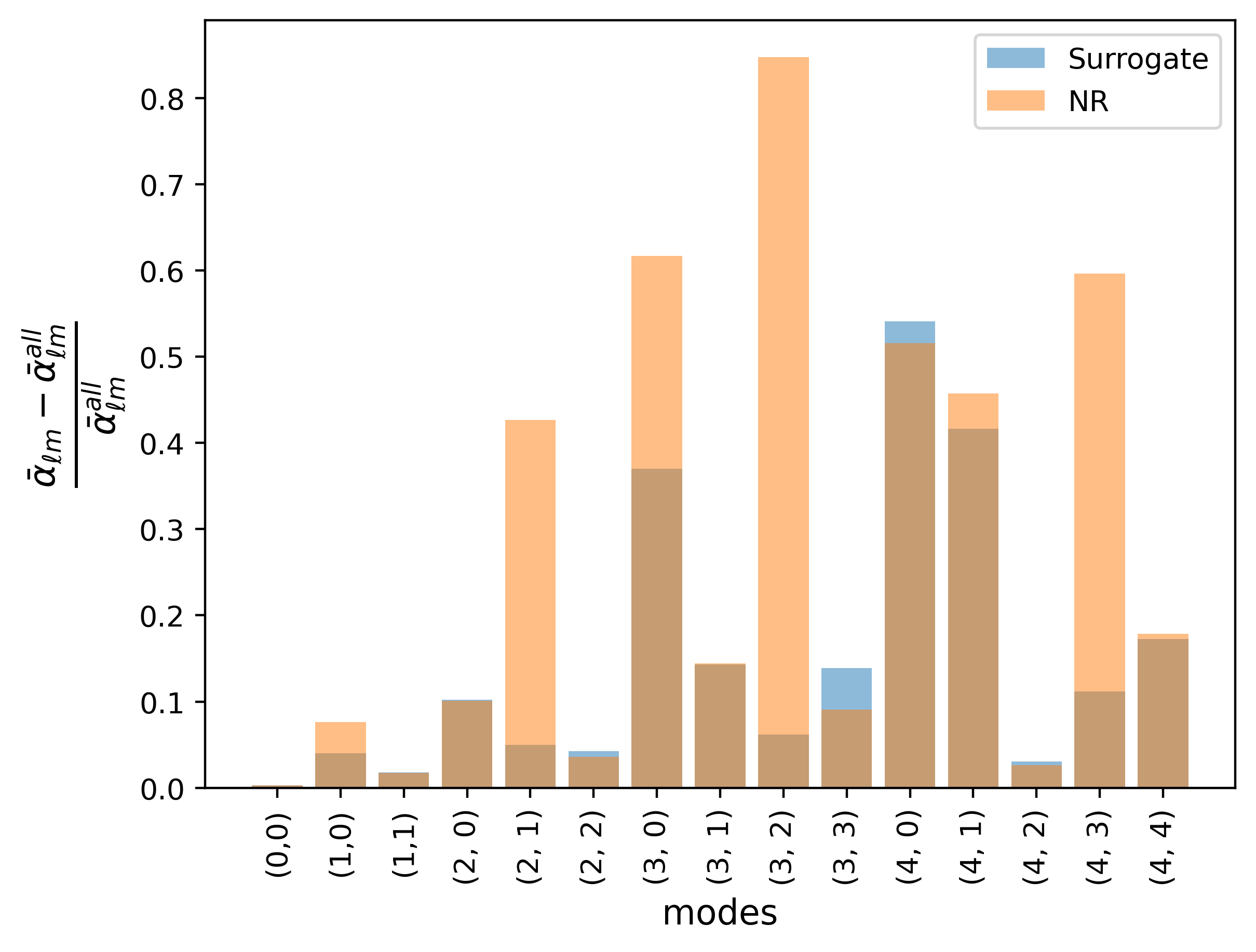}
	\caption{Time-integrated, relative error for $\alpha_{\ell m}$ containing all strain modes compared to neglecting high-mismatch modes for NR and \Surr{}.}
	\label{fig:alphas_rel}
\end{figure}
The Wigner-$3j$ symbols appearing in equation~\eqref{equ:alphas} single out individual terms $h_{\ell_1, m_1}h_{\ell_2, m_2}$\footnote{We suppress time derivative and conjugation symbols for ease of notation.} which contribute to each $\alpha_{\ell m}$. These coefficients, in turn, determine physical quantities such as the kick vector. 
According to equations~\eqref{equ:v1}-\eqref{equ:v3}, the coefficients relevant to the kick are $\alpha_{1,\pm1}$ and $\alpha_{1,0}$. Among all the modes combining to pairs within these coefficients, the most relevant mode combinations involve at least one dominant strain mode, $h_{2,\pm2}$. For the velocity components, such a mixing appears exclusively in $\alpha_{1,0}$, where it is observed that the term proportional to $h_{2,2}h_{2,2}$ cancels the term containing $h_{2,-2}h_{2,-2}$. Therefore, the most significant contributions to the kick components arise from the mixing of $h_{2,\pm1}$ modes with $h_{2,\pm2}$ in $\alpha_{1,\pm1}$\footnote{For $\alpha_{1\pm1}$, one finds also terms such as $h_{2,\pm2}h_{3,\pm1}$ and $h_{2,\pm2}h_{3,\pm1}$. However, the $\ell=2$ modes dominate over $\ell=3$.}. Contrarily, for $\alpha_{1,0}$ there is no non-vanishing contribution combining two $\ell=2$ strain modes, which demonstrates analytically the comparatively small $z$-component of the kick in non-precessing events. Naturally, the latter statement is not true in the precessing case as more power is distributed across subdominant modes, while $h_{2,2}$ still remains dominant.

Taking into account $\ell>2$ modes, for $\alpha_{1,\pm1}$, the strain mode $h_{2,2}$ also appears in combinations such as $h_{2,\pm2}h_{3,\pm1}$ and $h_{2,\pm2}h_{3,\pm3}$, which cancel in the sum~\eqref{equ:alphas}. Furthermore, when combining $\alpha_{1,1}$ and $\alpha_{1,-1}$ in the $v_x$ and $v_y$ velocity components, one of the components receives a larger contribution due to the cancellation of $h_{2,\pm2}h_{2,\pm1}$ terms in the sum $\alpha_{1,1} \pm \alpha_{1,-1}$. 

Using the above arguments, it follows that the dominant contribution in the $\alpha$-components relevant to the kick and proportional to $h_{2,\pm1}\notin H_{\text{sub}}$ are only marginally affected by the exclusion of subdominant modes $H_{\text{sub}}$; therefore, neither magnitude nor direction changes significantly when disabling $H_{\text{sub}}$ in the computation. A similar conclusion applies also to the final mass, which is determined by $\alpha_{0,0}$ and the strain mode mixing contained therein. 

The qualitative analysis of the mode contributions to the kick is quantified in Figure~\ref{fig:alphas_rel}. It displays the relative error of the $\alpha$-coefficients for NR and \Surr{} for $H_{\text{sub}}$ included versus excluded\footnote{\EOB{} and \Phen{} are excluded from the presentation as their $\alpha$-coefficients do not show a noticeable change when $H_{\text{sub}}$ is omitted. In these models, every mode in $H_{\text{sub}}$ is already vanishing.}. The plot clearly illustrates that the alteration in the $\alpha$-coefficients relevant for the kick velocity is minimal. The more pronounced difference in NR is explained by its expanded mode content ($\ell\leq 8$) compared to \Surr{} ($\ell\leq 4$), influencing a greater number of mode mixing terms in the summation of $\alpha_{1\pm1}$ when $H_{\text{sub}}$ is disregarded.

In conclusion, the fundamental limit of the kick velocity in assessing subdominant modes results from the restricted mixing appearing in $\alpha_{1m}$. Based on the Wigner-$3j$ symbols in~\eqref{equ:alphas}, the allowed coupling of modes is restricted to $|\ell_1-\ell_2|\leq\ell$ and, correspondingly, the dominant $h_{2,2}$ mode does not combine with any higher modes. This, however, changes when computing the memory.

\subsubsection{Memory Components}
The linear and non-linear GW memory induced by the waveforms is computed and decomposed in equation~\eqref{eq:mem_total_def}.
\begin{figure}
   \includegraphics[width=0.9\columnwidth]{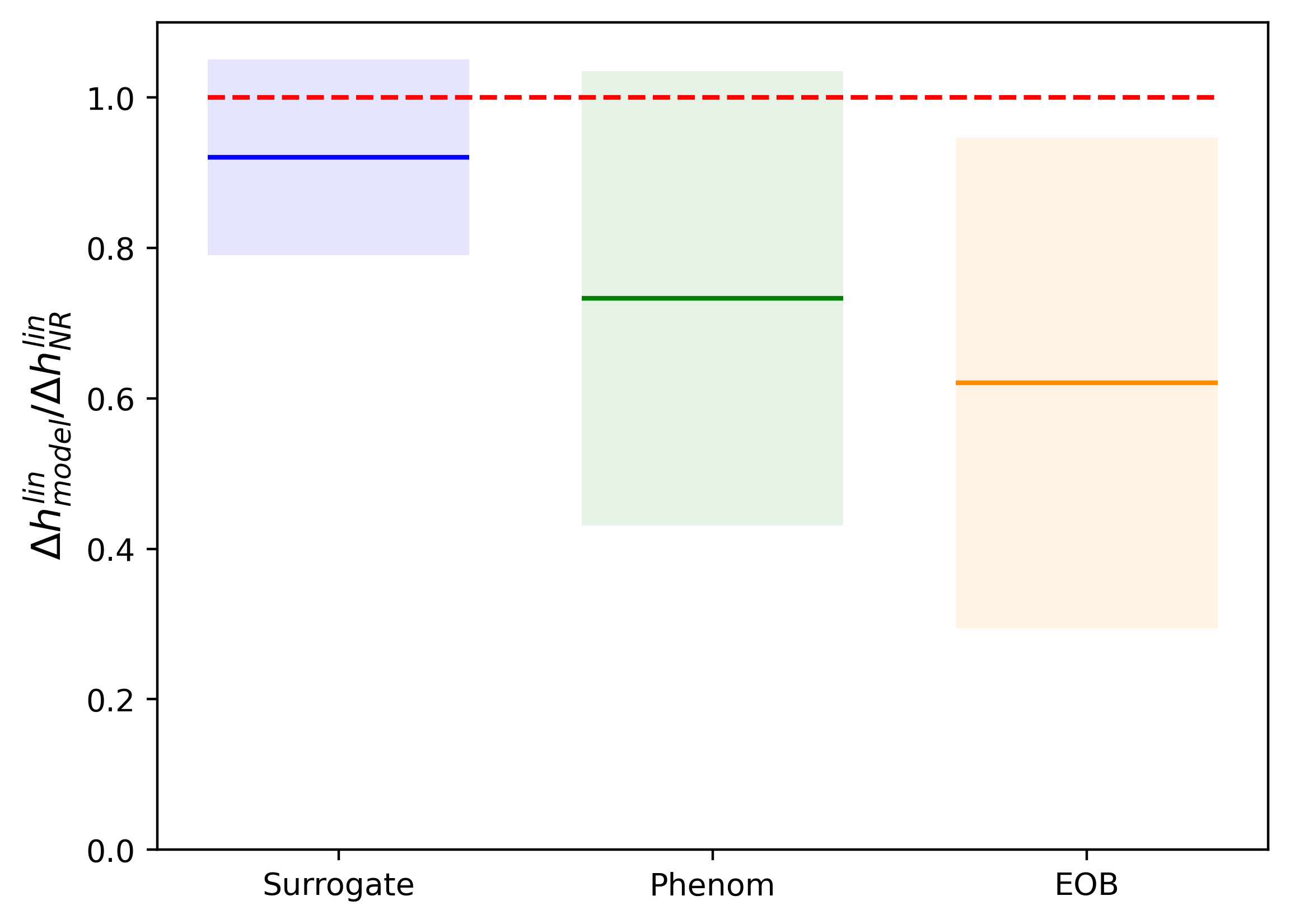}    
\includegraphics[width=0.9\columnwidth]{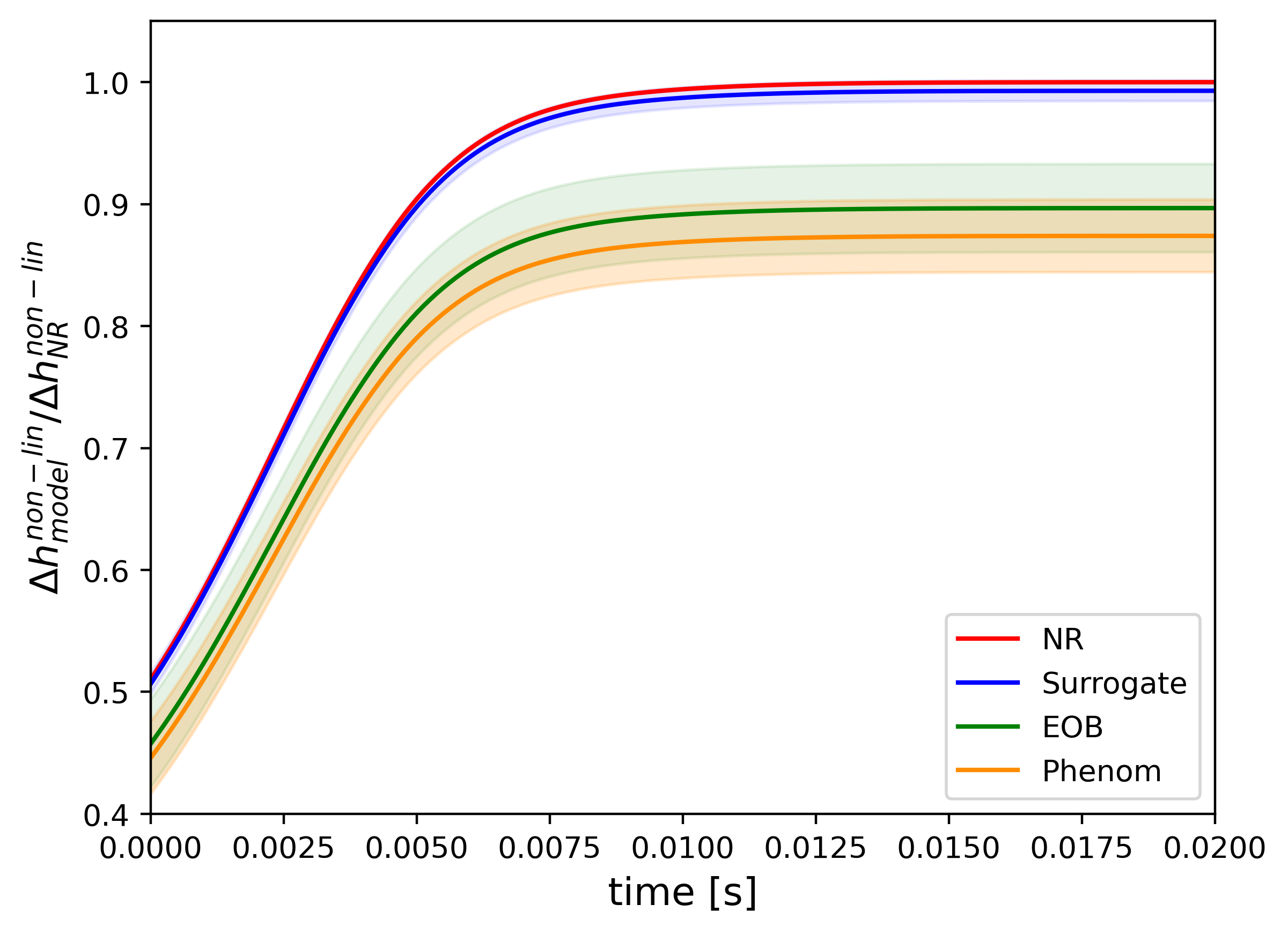}   
     \caption{Deviation of waveform models' linear (top) and non-linear (bottom) memory with respect to NR. Corresponding $2\sigma$-intervals are colorized. The non-linear memory is plotted as a time series. Both memory contributions are normalized to the NR value and averaged over all events.}
       \label{fig:memory_error}
\end{figure}
Figure~\ref{fig:memory_error} illustrates the relative errors for both linear and non-linear contributions to the memory in non-precessing events, considering the full mode content. In the upper plot, the total linear memory is normalized to the NR contribution for each waveform model. The colorized regions indicate the $2\sigma$ interval of the average over all events. The solid lines represent the mean values. Similar quantities are presented in the bottom plot for the non-linear memory, depicted as a time series to observe the error accumulation as the non-linear memory builds up. The time $t = 0$ corresponds to the merger time, and the normalization is performed with respect to the time-integrated non-linear NR memory.

In terms of both linear and non-linear memory, \EOB{} and \Phen{} exhibit notably poorer performance compared to the \Surr{} approximant. These two models consistently produce smaller memory values. In contrast, \Surr{} performs well, particularly in the case of the dominant non-linear GW memory. For \EOB{} and \Phen{}, the NR values fall outside the $2\sigma$ range. \EOB{} also shows a similar trend for the linear part, suggesting systematic deficiencies regarding the memory in these models. 

In general, the linear memory is highly non-trivial to model. Equation~\eqref{eq:lin_mem_def} was derived under the assumption that the waveform (and modes) is well-defined throughout time, which is a crude approximation. A more exact formulation of the linear memory is given in~\cite{Mitman:2020bjf}, which would require the extraction of certain Newman-Penrose scalars not provided for the majority of events. In addition to the considered approximation, the linear part is many orders of magnitude smaller than the non-linear part, making it more prone to numerical errors in the models. Consequently, the linear memory exhibits large statistical fluctuations.

As before, we now compute the memory without the use of the subdominant modes $H_{\text{sub}}$. We find that, for the linear memory component, the change of mode content, similar to the kick velocity, results only in a marginal change of the models individual contributions. The reason is found in~\eqref{eq:lin_mem_def}, where the defined linear memory is solely determined by the remnant velocity. This dependency on the kick is a result of the approximation used in the derivation of the balance laws as we state them, i.e., \eqref{equ:fullDimBL}, where the Newman-Penrose scalars originally appearing in the balance balance laws can be replaced by functions of the remnant velocity in certain limits\footnote{For a detailed discussion on the relevant transformations and physical assumptions we refer to \cite{Ashtekar:2019viz}.}. Consequently, as the linear memory is determined by the kick, the discussion regarding the strain modes entering its computation follows similar arguments as for the kick in the previous subsection. 

\begin{figure}
\includegraphics[width=1.\columnwidth]{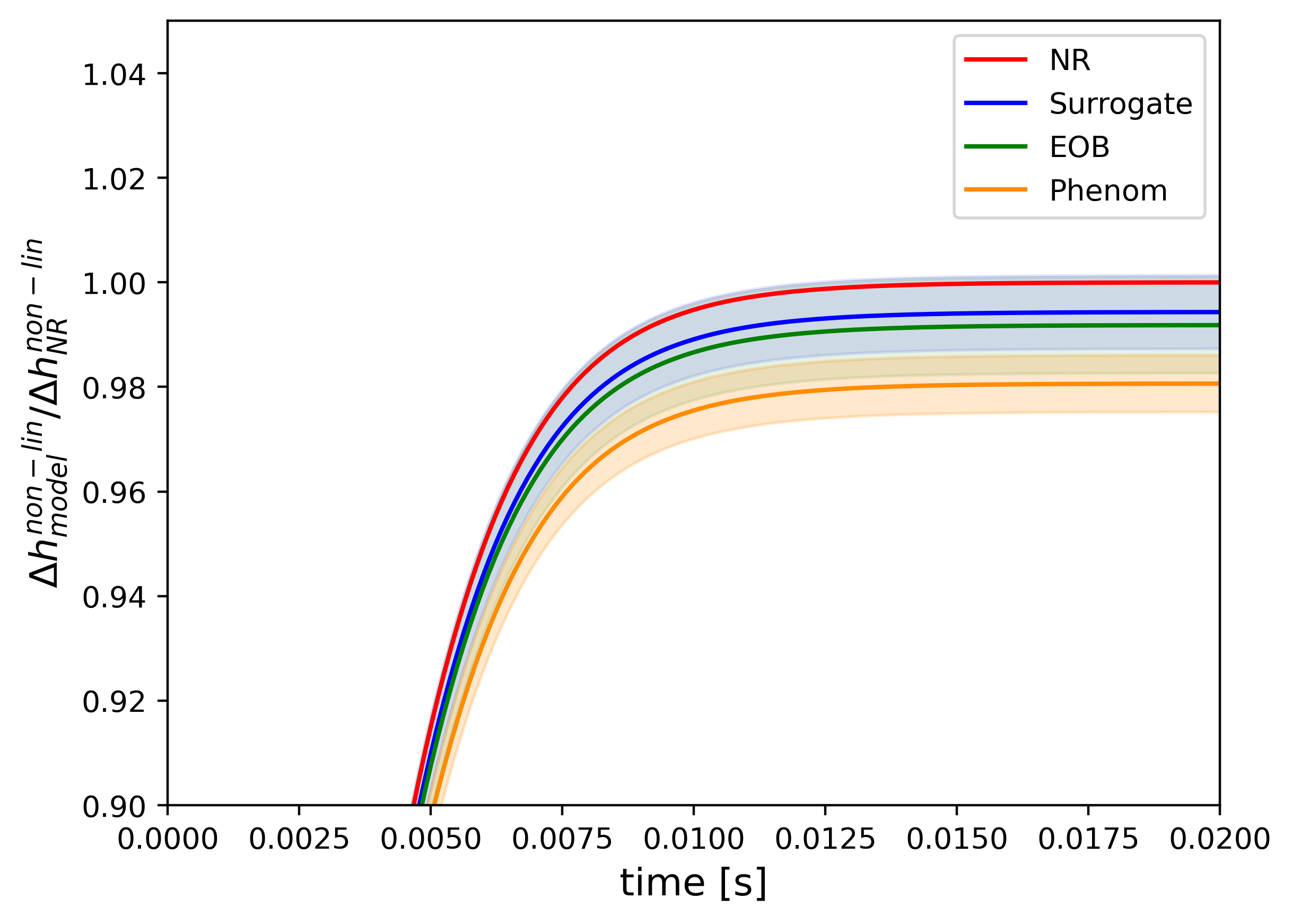}  
     \caption{Non-linear memory as depicted in Figure~\ref{fig:memory_error}, here computed without high-mismatch modes, i.e., excluding $H_{\text{sub}}$.}
    \label{fig:memory_error_II}
\end{figure}

A completely different picture is drawn by the non-linear memory when disabling the set of modes $H_{\text{sub}}$. As depicted in Figure~\ref{fig:memory_error_II}, the relative errors throughout models are approaching each other when the mode content is reduced. While the absolute values for NR and \Surr{} have decreased by $\mathcal{O}(10)$ percent, \EOB{} and \Phen{} have virtually not changed. Consequently, after excluding $H_{\text{sub}}$, the non-linear memory contribution of all considered approximants is roughly equal, and, by the definition of the balance laws, so is their energy flux. This statement is independent of the alignment procedure as the non-linear memory is not sensitive to the chosen reference phase\footnote{This is, in principle, also true for both non-precessing and precessing events, where for the latter spins have to be taken into consideration as well.}. Unlike the non-linear memory, the kick velocity, and thus the linear memory, is sensitive to the alignment procedure. Therefore, some of the discrepancies shown in the top of Figure~\ref{fig:memory_error} may be attributed to suboptimal alignment. 

As anticipated by previous literature (e.g. \cite{Borchers_2023} and compare to findings in \cite{Khera:2020mcz}), \EOB{} and \Phen{} models do not reproduce the full information content with their generated strain modes. 
Comparing Figures~\ref{fig:memory_error} and \ref{fig:memory_error_II}, it becomes apparent that the modes contained in $H_{\text{sub}}$ roughly constitute the discrepancy in (memory) information content transmitted through the waveforms of \EOB{} and \Phen{} with respect to NR (and \Surr{}). 
To explicitly demonstrate this, we once again turn to the analysis of strain mode mixing in the memory components:
According to equation~\eqref{eq:nonlin_mem_def}, the total non-linear memory contains all $\alpha$-coefficients with $\ell>1$, where $\alpha_{2,0}$ dominates the sum of coefficients. This is a consequence of the mode mixing present in the $\alpha$-coefficients\footnote{Physically, this results from the fact that the majority of energy flux away from the binary is transported by isotropic gravitational radiation. The azimuthally symmetric flux is sourced by $h_{2,\pm2}$. Its manifestation in terms of memory is carried by $h_{2,0}$.}. As we already elaborated, two dominant strain modes $h_{2,2}$, $h_{2,-2}$ can only mix for $\alpha$-coefficients such as $\alpha_{\ell,0}$. However, the $h_{2,2}h_{2,2}$ and $h_{2,-2}h_{2,-2}$ terms cancel for all $\ell$, except for $\alpha_{2,0}$ and $\alpha_{4,0}$. For $\alpha_{\ell, m}$ with $\ell>4$ such couplings are forbidden by virtue of the Wigner-$3j$ symbols in~\eqref{equ:alphas}. For $\alpha_{4,0}$ the prefactor in~\eqref{equ:alphas} is of order $\mathcal{O}(10^{-2})$, suppressing its contribution to the full non-linear part. Consequently, on average, $\alpha_{2,0}$ accounts for roughly $96\%$ of the non-linear memory.

Naturally, the majority of $\alpha_{2,0}$ is due to the term including two dominant modes. However, there is a significant part of the $\alpha_{2,0}$ component as well as the contributions from all other $\alpha$-coefficients that are due to subdominant modes coupling to $h_{2,\pm2}$. As many more $\alpha$'s enter the non-linear memory and for $\ell>1$ more mixing of the strain modes is permitted by the Wigner-$3j$ symbols, the impact of subdominant modes is non-negligible (in contrast to what was observed for the kick velocity) and grows with the number of strain modes included in the computation of the non-linear memory. The difference in the non-linear memory between Figures~\ref{fig:memory_error} and~\ref{fig:memory_error_II} is, hence, largely due to the coupling of $h_{2,2}$ with subdominant modes in $\alpha$'s with $\ell>1$. Switching of the (for \EOB{} and \Phen{}) missing subdominant modes in NR and \Surr{}, reduces the memory mainly to the $h_{2,\pm2}h_{2,\pm2}$ couplings that, due to the well modeled dominant strain mode throughout approximants, are roughly equal to \EOB{} and \Phen{} as indicated by Figure~\ref{fig:memory_error_II}.

\begin{figure}
\includegraphics[width=1.\columnwidth]{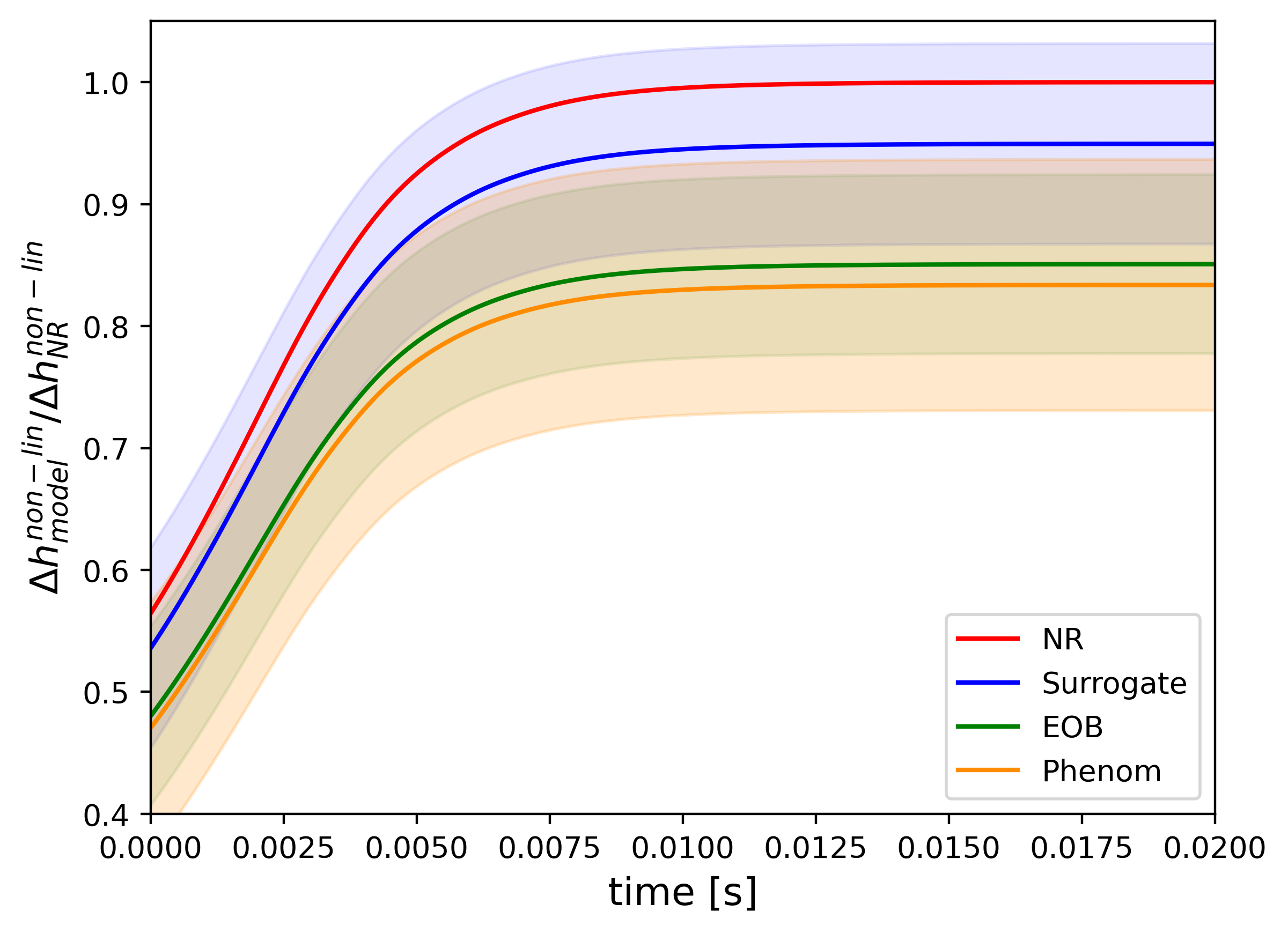}  
\includegraphics[width=1.\columnwidth]{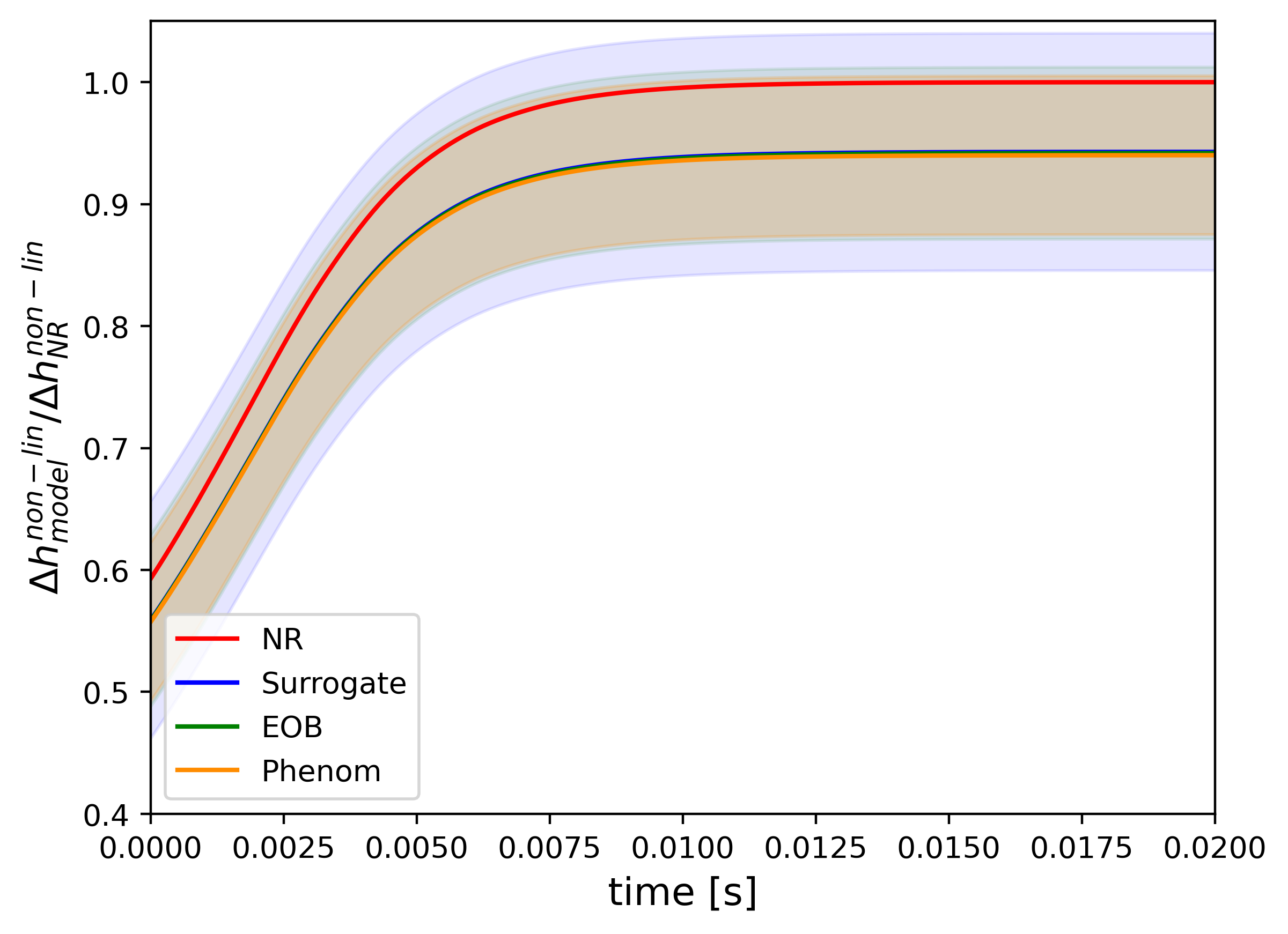}   
     \caption{Non-linear memory as depicted in Figure~\ref{fig:memory_error}, here computed with (top) and without (bottom) high-mismatch modes, i.e., excluding $H_{\text{sub}}$, for precessing events.}
       \label{fig:memory_error_III}
\end{figure}
Similar results are obtained in the precessing case, as shown in Figure~\ref{fig:memory_error_III}. In the top plot, the average relative errors for the full mode content are displayed as time series. It is observed that \EOB{} and \Phen{} exhibit similar deviations from NR as in the non-precessing case. Additionally, \Surr{} obtains considerably larger deviations compared to non-precessing events and displays a large standard deviation.
The latter persists when $H_{\text{sub}}$ is excluded, as shown in the bottom plot of Figure~\ref{fig:memory_error_III}. Nonetheless, for the non-linear memory, it still performs better than \EOB{} and \Phen{} including the full mode content. When the modes in $H_{\text{sub}}$ are disabled, \Surr{}, \EOB{} and \Phen{} exhibit identical memory contributions. Notably here, \Surr{} has the largest $2\sigma$ interval. 

Although the non-linear memory contributions of the approximants closely resemble each other, we do not observe a similar convergence to NR as in Figure~\ref{fig:memory_error_II}. Instead, all models settle on average on $90\%$ of NR's total non-linear memory. This indicates that for the precessing case, \Surr{}, \EOB{} and \Phen{} differ mainly by modes included in $H_{\text{sub}}$, however, to reproduce NR all approximants are evidently missing information beyond what is contained in $H_{\text{sub}}$ as indicated by the gap between approximants and NR in the bottom plot of Figure~\ref{fig:memory_error_III}. Additional discrepancies might manifest in the remnant velocity for precessing events. However, since the alignment procedure for precessing events is not effective as outlined in subsection~\ref{subsec:alignment}, we cannot assess the kick approximation of the waveform models for precessing events in a meaningful way.

Leveraging the complete parameter space detailed in~\ref{sec:alignmentandparas}, we examine whether the conclusions concerning the memory are influenced by the choice of NR events. By considering the supplementary parameter space beyond the \SXS{} data, we verify that the results outlined earlier remain consistent for both precessing and non-precessing events. The relations between the memory contributions of \Surr{}, \EOB{}, and \Phen{} are validated for these instances, affirming that our observations are not affected by potential biases in the selection of extrinsic parameters.

Concluding our findings, we observe that neither the kick nor the memory in this form are able to assess a models subdominant waveform contribution. They are, however, excellent probes for the dominant $h_{2,\pm2}$ mode and for the mode content in general. The non-linear memory serves as a valuable tool to distinctly highlight the deficiencies in the mode content offered by \EOB{} and \Phen{} models considered in this study. However, memory and kick as computed in this work indicate that the missing modes are not the only source of discrepancies of \EOB{} and \Phen{} with respect to NR (or \Surr{}). Our results imply that disparities emerge even at the level of subdominant modes common to all models. Specifically, we highlight the mixing of $h_{2,\pm1}$ with $h_{2,2}$ in the $\alpha$ coefficients relevant to the kick, which significantly contributes to the observed dissimilarities in kick magnitude and its direction as illustrated in Figure~\ref{fig:error_kick_NR_np}.

\section{Discussion}
\label{sec:discussion}
We conducted an extensive comparison of gravitational waveform models based on the calculation of the remnant's kick velocity and the gravitational memory for BBH mergers. Four state-of-the-art waveform models were considered (cf.  Table~\ref{table:1}), each representing a distinct approach in efficiently generating a gravitational waveform for a given BBH configuration. We reviewed their main features, dissimilarities, and conventions. Moreover, we discussed issues in the commonly used waveform alignment procedures, pointing out difficulties in the meaningful distinction between alignment residuals and intrinsic waveform dissimilarities. After applying a well-tested alignment strategy, we utilized so-called balance laws to compute the kick velocity and the memory. In order to minimize a potential selection bias, we extended our analysis beyond the catalogued \SXS{} BBH events and additionally considered randomly generated events in parameter space. We compared the computed kick velocities and memory terms with respect to reference models. In the case of catalogued events we chose NR waveforms as reference, while \Surr{} was used for events not contained in the \SXS{} database. 

Our results demonstrate inconsistencies between the selected approximants and the reference model regarding both the computed kick velocity and inferred memory. This corroborates earlier findings in the literature~\cite{Borchers_2023}. We found that the \Surr{} models considered here clearly outperform \EOB{} and \Phen{} waveforms. It was further demonstrated that \EOB{} and \Phen{} models do not accurately determine the gravitational wave memory due to the reduced subdominant mode content. This is true for both precessing and non-precessing events, and in good approximation independent of the alignment. For precessing events, the non-linear memory of all approximants, including \Surr{}, display inconsistencies with respect to the \SXS{} waveform memory as shown in Figure~\ref{fig:memory_error_III}.
On the other hand, compared to the gravitational wave memory, our results show that kick velocity is more strongly influenced by the dominant strain modes $h_{2,\pm2}$ and less sensitive to subdominant modes, both in amplitude and direction. Removing certain subdominant modes did not yield a significant deviation in the computed kick velocity vector.

We support our findings by analytical considerations based on strain mode mixing in the spherical harmonic decomposition of $(|\dot h|^2)_{\ell,m}$, as described by the $\alpha$-coefficients of equation~\eqref{equ:alphas}. Notice that the way the modes mix is ultimately determined by the balance laws~\eqref{equ:BLinModes} and that with the help of the $\alpha$-coefficients one can express physical quantities of the BBH merger, such as radiated energy or remnant mass, kick velocity, and gravitational memory\footnote{When calibrating our numerical pipeline, we test all computations of physical quantities against the metadata contained in the \SXS{} database. For instance, the kick velocity magnitudes computed from~\eqref{equ:kick} were compared against the values in the \SXS{} data. The substantial agreement found in these comparisons validate the applicability of the balance laws.}. Another use of the $\alpha$-coefficients, which we discussed in~\ref{sec:analysis}, is to explain the relationship between the azimuthal symmetry of the energy flux away from the binary and the dominant memory contribution residing in the (azimuthally independent) $h_{2,0}$ mode. 

Generally, our findings indicate that while the kick velocity as defined in section~\ref{sec:theory} is an excellent estimator of the quality of $h_{2,\pm2}$ mode and highly sensitive to the alignment procedure, the GW memory serves as a good indicator for differences in the mode content of the waveforms, both for precessing and non-precessing events. However, towards individual subdominant modes, or sets of those, we find that quantities like kick and memory in their original form exhibit very low sensitivity. 

In fact, one key conclusion which can be drawn from our analysis is that kick velocity and gravitational memory are not suitable as high-precision metrics for modes with $\ell>2$. In light of this realization, we propose a new three-step strategy for future waveform comparisons:
Firstly, for each approximant, waveforms are generated and set on equal footing regarding the spatial and temporal frame (for instance via the \LALSuite). After the application of a suitable alignment scheme, such as for instance the one applied in this article, residual ambiguities in the waveforms are resolved. For each approximant, the waveforms and their (provided) mode decomposition are now ready for the computation of physical quantities. Thus, we compute the $\alpha$-coefficients using equation \eqref{equ:alphas} in combination with the aligned waveforms. 

The first step is repeated for different strain mode contents, i.e., in each iteration, the set of strain modes included in the computation of the $\alpha$-coefficients changes depending on which section of the mode content shall be tested. For instance, being interested in the overall impact of $\ell=3,4$ modes on the precision of the waveform, the first step yields a set $\{\alpha_{\ell,m}^{\text{approx.}}(h_{\ell,m}), \alpha_{\ell,m}^{\text{approx.}}(h_{\ell,m}|\ell\neq 3),\alpha_{\ell,m}^{\text{approx.}}(h_{\ell,m}|\ell\neq 4)\}$ for each approximant\footnote{We can even be more precise here and additionally specify the $m$ of an individual mode that we wish to analyze. Note, however, that it is crucial to remove both $h_{\ell,\pm m}$ in this case.}.

In the second step, we use $\alpha_{\ell,m}^{\text{approx.}}(h_{\ell,m})$ to compute the remnant velocity vector. Using its overall amplitude and direction, we test and compare the waveforms' dominant $h_{2,\pm2}$ and subdominant $h_{2,\pm1}$ mode under the assumption that an preceding alignment procedure minimized the mismatch and only intrinsic dissimilarities between waveforms stemming form different approximates remain\footnote{This step reproduces previous investigations such as for instance \cite{Borchers_2023}}. The kick tests both of the latter modes simultaneously as it is dominated by the strain mode $h_{2,\pm1}h_{2,\pm2}$. For a proper distinction of the origin of dissimilarities, the $h_{2,2}$-mismatch and kick can be used in synergy.

Provided that the dominant $h_{2,\pm2}$ mode is modeled sufficiently accurate, the third step of the proposed strategy comprises a comparison of the computed $\alpha$'s between approximants. Here, we one can proceed twofold, i) the $\alpha$-coefficients are compared in accumulated form as the non-linear memory, or ii) $\alpha$-coefficients are compared individually. In the gravitational wave memory, the effect of individual subdominant strain modes accumulates and, thus, can be quantified despite the mode in question being highly subdominant. Concretely, the memory contributions computed using the different sets of $\alpha$'s within an approximant are compared with each other to gauge the impact of individual modes for a given approximant. The corresponding errors are then compared to the same quantities computed for \NR{} waveforms. Shortcomings in certain modes are exposed if excluding these yields a larger impact for the \NR{} than for the approximant's memory contribution. The shortcomings can be addressed by phenomenological corrections in the waveform or individual strain modes.

In case of a successful implementation of i), we can further improve precision of the analysis and proceed with ii): Each $\alpha_{\ell,m}$ contains different subdominant modes coupling to $h_{2,\pm2}$ (or the next-to-dominant mode, e.g, $h_{2,\pm1}$) which can be tested by simply comparing the corresponding $\alpha$-coefficients of \NR{} and the approximants against each other. Here, the previous removal of subdominant modes isolates the impact of a given mode in question within a specific $\alpha_{\ell,m}$. For instance, removing all strain modes with $m=0$ in the computation of the $\alpha$-coefficients will isolate the $h_{2,\pm2}h_{4,\pm4}$ mixing in $\alpha_{2,2}$ as the only dominant mode mixing. Hence, with the previous steps executed, comparing the so computed $\alpha_{2,2}$ between \NR{} and the approximants will indicate potential discrepancies in the $h_{4,\pm4}$ mode. Note, however, that only strain modes coupling directly to $h_{2,\pm2}$ in the $\alpha$-coefficients can be tested in this way. For modes that only couple to subdominant modes in the presence of another dominant-mode-coupling, the metrics of comparison computed according to the above strategy are statistically not significant when considering an ensemble of events. On the other hand, the advantage of ii) over i) resides in a more precise mode testing. That is, while the memory influenced by many subdominant modes simultaneously, its individual components depend only a small number of couplings to $h_{2,\pm2}$. Testing an individual mode using the memory, the accumulation of mode impact in the sum over $\alpha$'s might pick up discrepancies in other subdominant modes coupling to the one in question. 

The third step of the prescribed strategy targets predominantly the mode content with $\ell>2$. As pointed out in section \ref{sec:analysis}, these modes exhibit particularly large mismatch or are trivial for \EOB{} and \Phen{}. Thus, the latter step is especially relevant for updated models of \EOB{} and \Phen{} (e.g.~\cite{pompili2023laying, ramosbuades2023seobnrv5phm} and future \Phen{} models) that contain more modes than the versions tested in this work.  

Physically speaking, testing individual $\alpha$-coefficients corresponds to analysing the anisotropic non-linear memory, or equivalently, the anisotropic energy flux radiated to infinity from the binary system. Naturally, the anisotropic energy flux is highly configuration-dependent and subdominant. It nevertheless constitutes a powerful tool in high-precision tests of subdominant waveform contributions, especially for precessing events.

The advantage of assessing a waveform's quality by systematically deleting subdominant modes from the computation of physical quantities, as opposed to a mode-by-mode mismatch analysis as the one depicted in Figure~\ref{fig:mismatch_NR_np}, is that no fit to the corresponding NR counterpart is needed. Such a fitting procedure would defeat the purpose of the waveform approximants, since they are supposed to produce accurate waveforms in a more time-efficient manner than NR.

We leave an implementation of the outlined strategy for future work. Here, we have established the groundwork for these forthcoming systematic studies by developing codes, methods, and guidelines that are applicable to the important task of comparing and assessing the accuracy and reliability of waveform models. 


\section*{Acknowledgements}
LH would like to acknowledge financial support from the European Research Council (ERC) under the European Unions Horizon 2020 research and innovation programme grant agreement No 801781. LH further acknowledges support from the Deutsche Forschungsgemeinschaft (DFG, German Research Foundation) under Germany’s Excellence Strategy EXC 2181/1 - 390900948 (the Heidelberg STRUCTURES Excellence Cluster). The authors thank the Heidelberg STRUCTURES Excellence Cluster for financial support. The authors acknowledge support by the state of Baden-Württemberg, Germany, through bwHPC. The authors thank Cecilio Garcia-Quiros, Keefe Mitman and Nils Deppe for fruitful discussions.


\appendix

\section{Numerical artifacts}
\label{sec:numerical_artefacts}
\begin{figure}[htb!] 
	\centering
	\includegraphics[width=0.99\linewidth]{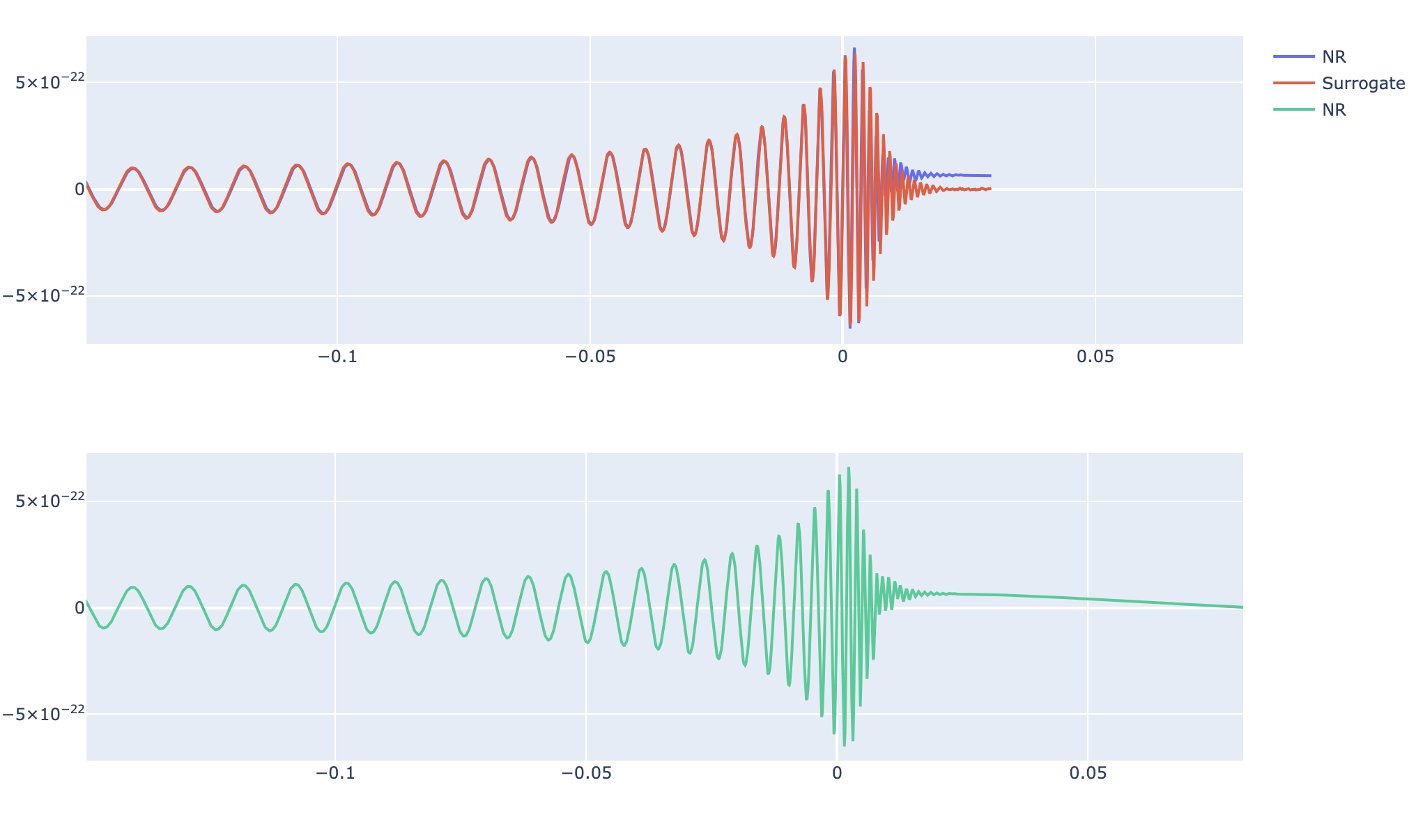}
	\caption{$(4,4)$ waveform mode for a non-precessing event SXS:BBH:$0155$. Displayed are the truncated and aligned waveforms (above) for NR and \Surr{} as well as the ``raw'' full waveform for NR.}
	\label{fig:artefact}
\end{figure}

Numerical artifacts can significantly influence the outcome of the mismatch function discussed in section~\ref{sec:alignmentandparas}. The morphology with which artefacts can appear varies. One of the most prominent examples is displayed in Figure~\ref{fig:artefact}.
Here, the waveform mode carries a tail, which shall not be confused with the gravitational memory. The latter is not contained in the waveforms displayed. Instead, the non-trivial extent of the waveform after the ring down results from numerical artifacts and is to be disregarded as non-physical. Since the effect is rather small and effects only subdominant modes, we ignored it in the main analysis. In future works tackling the precision of higher strain modes, this issue has to be addressed.

\bibliographystyle{apsrev4-2}
\bibliography{reference.bib}

\end{document}